\pdfoutput=1

\documentclass[11pt,twoside,a4paper,cmspaper,final,collab]{cms-tdr}
\def\svnVersion{340318}\def\cmsCernNoTag{CERN-PH-EP/2015-149}\def\cmsCernDate{\today}\def\cmsMessage{Published in Physical Review D as \href{http://dx.doi.org/10.1103/PhysRevD.93.032005}{\doi{10.1103/PhysRevD.93.032005.}}}

\begin{document}\cmsNoteHeader{EXO-12-043}

\hyphenation{had-ron-i-za-tion}
\hyphenation{cal-or-i-me-ter}
\hyphenation{de-vices}
\RCS$HeadURL: svn+ssh://svn.cern.ch/reps/tdr2/papers/EXO-12-043/trunk/EXO-12-043.tex $
\RCS$Id: EXO-12-043.tex 311495 2015-11-22 15:27:24Z alverson $
\newlength\cmsFigWidth
\ifthenelse{\boolean{cms@external}}{\setlength\cmsFigWidth{0.85\columnwidth}}{\setlength\cmsFigWidth{0.4\textwidth}}
\ifthenelse{\boolean{cms@external}}{\providecommand{\cmsLeft}{top}}{\providecommand{\cmsLeft}{left}}
\ifthenelse{\boolean{cms@external}}{\providecommand{\cmsRight}{bottom}}{\providecommand{\cmsRight}{right}}
\ifthenelse{\boolean{cms@external}}{\providecommand{\cmsTopLeft}{top}}{\providecommand{\cmsTopLeft}{top left}}
\ifthenelse{\boolean{cms@external}}{\providecommand{\cmsTopRight}{middle}}{\providecommand{\cmsTopRight}{top right}}
\ifthenelse{\boolean{cms@external}}{\providecommand{\cmsBottom}{bottom}}{\providecommand{\cmsBottom}{bottom}}
\ifthenelse{\boolean{cms@external}}{\providecommand{\CL}{C.L.\xspace}}{\providecommand{\CL}{CL\xspace}}
\ifthenelse{\boolean{cms@external}}{\providecommand{\NA}{\ensuremath{\cdots}}\xspace}{\providecommand{\NA}{---\xspace}}
\providecommand{\CLS}{CL$_{S}$\xspace}
\newcolumntype{.}{D{.}{.}{-1}}

\newcommand{\ST}{\ensuremath{S_{\mathrm{T}}}\xspace}

\cmsNoteHeader{EXO-12-043}
\title{Search for single production of scalar leptoquarks in proton-proton collisions at \texorpdfstring{$\sqrt{s} = 8$ TeV}{sqrt(s) = 8 TeV}}

\date{\today}

\abstract{
        A search is presented for the production of both first- and second- generation scalar leptoquarks with a final state of either two electrons and one jet or two muons and one jet.  The search is based on a data sample of proton-proton collisions at center-of-mass energy $\sqrt{s}=8$\TeV recorded with the CMS detector and corresponding to an integrated luminosity of 19.6\fbinv.  Upper limits are set on both the first- and second- generation leptoquark production cross sections as functions of the leptoquark mass and the leptoquark couplings to a lepton and a quark.  Results are compared with theoretical predictions to obtain lower limits on the leptoquark mass.  At 95\% confidence level, single production of first-generation leptoquarks with a coupling and branching fraction of 1.0 is excluded for masses below 1730\GeV, and second-generation leptoquarks with a coupling and branching fraction of 1.0 is excluded for masses below 530\GeV.  These are the best overall limits on the production of first-generation leptoquarks to date.
}

\hypersetup{
pdfauthor={CMS Collaboration},
pdftitle={Search for single production of scalar leptoquarks in pp collisions at sqrt(s) = 8 TeV with the CMS Detector},
pdfsubject={CMS},
pdfkeywords={CMS, physics, software, computing}}

\maketitle

\section{Introduction}
\label{introduction}

Leptoquarks (LQ) are hypothetical color-triplet bosons with spin 0 (scalar LQ) or 1 (vector LQ), which are predicted by many extensions of the standard model (SM) of particle physics, such as Grand Unified Theories~\cite{PhysRevLett.32.438,Pati:1973uk,Pati:1974yy,Murayama:1991ah,Fritzsch:1974nn,Senjanovic:1982ex,PhysRevLett.65.2209,Frampton:1990hz}, technicolor schemes~\cite{Dimopoulos:1979es,Dimopoulos:1979sp,Farhi:1980xs}, and composite models~\cite{Schrempp:1984nj}. They carry fractional electric charge ($\pm1/3$ for LQs considered in this paper) and both baryon and lepton numbers and thus couple to a lepton and a quark.  Existing experimental limits on flavor changing neutral currents and other rare processes disfavor leptoquarks that couple to a quark and lepton of more than one SM generation~\cite{Buchmuller:1986iq,FCNC}.  A discussion of the phenomenology of LQs at the LHC can be found elsewhere~\cite{Belyaev:2005ew}.

The production and decay of LQs at proton-proton colliders are characterized by the mass of the LQ particle, $M_{\text{LQ}}$; its decay branching fraction into a charged lepton and a quark, usually denoted as $\beta$; and the Yukawa coupling $\lambda$ at the LQ-lepton-quark vertex. At hadron colliders, leptoquarks could be produced in pairs via gluon fusion and quark anti-quark annihilation, and singly via quark-gluon fusion. Pair production of LQs does not depend on $\lambda$, while single production does, and thus the sensitivity of single LQ searches depends on $\lambda$.  At lower masses, the cross sections for pair production are greater than those for single production.  Single production cross sections decrease more slowly with mass, exceeding pair production at an order of 1 \TeV for $\lambda$ = 0.6.

Several experiments have searched for LQs.  The H1 collaboration has produced limits on various singly produced LQ types: the one to which to compare this search is the LQ called $S^{R}_{0}$ in Ref.~\cite{Collaboration:2011qaa}, for which they place a limit at 500\GeV, assuming $\lambda=1.0$ and $\beta=1.0$.  The \DZERO collaboration has produced limits on singly produced scalar LQs of 274\GeV, again assuming $\lambda=1.0$ and $\beta=1.0$~\cite{Abazov:2006ej}.  Limits from pair production of leptoquarks exclude leptoquark masses below 1010\GeV for the first generation and 1080\GeV for the second generation, for $\beta=1.0$~\cite{PairProd}.

\begin{figure}[!h]
       \centering
       \includegraphics[scale=.45]{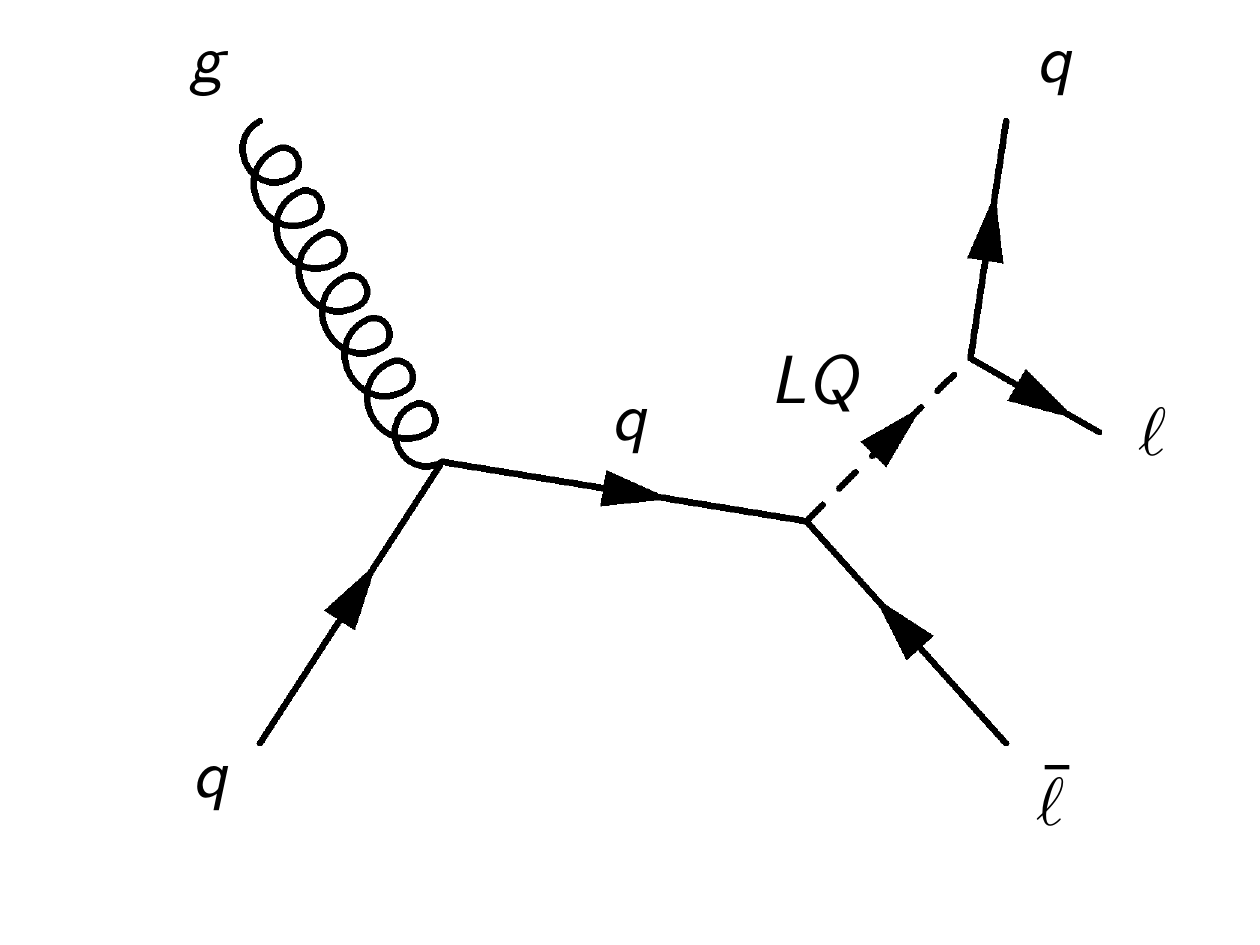}
       \caption{The $s$-channel resonant LQ production diagram.
	  \label{figapp:ResDiagram}}
\end{figure}

\begin{figure}[!h]
       \centering
       \includegraphics[scale=.45]{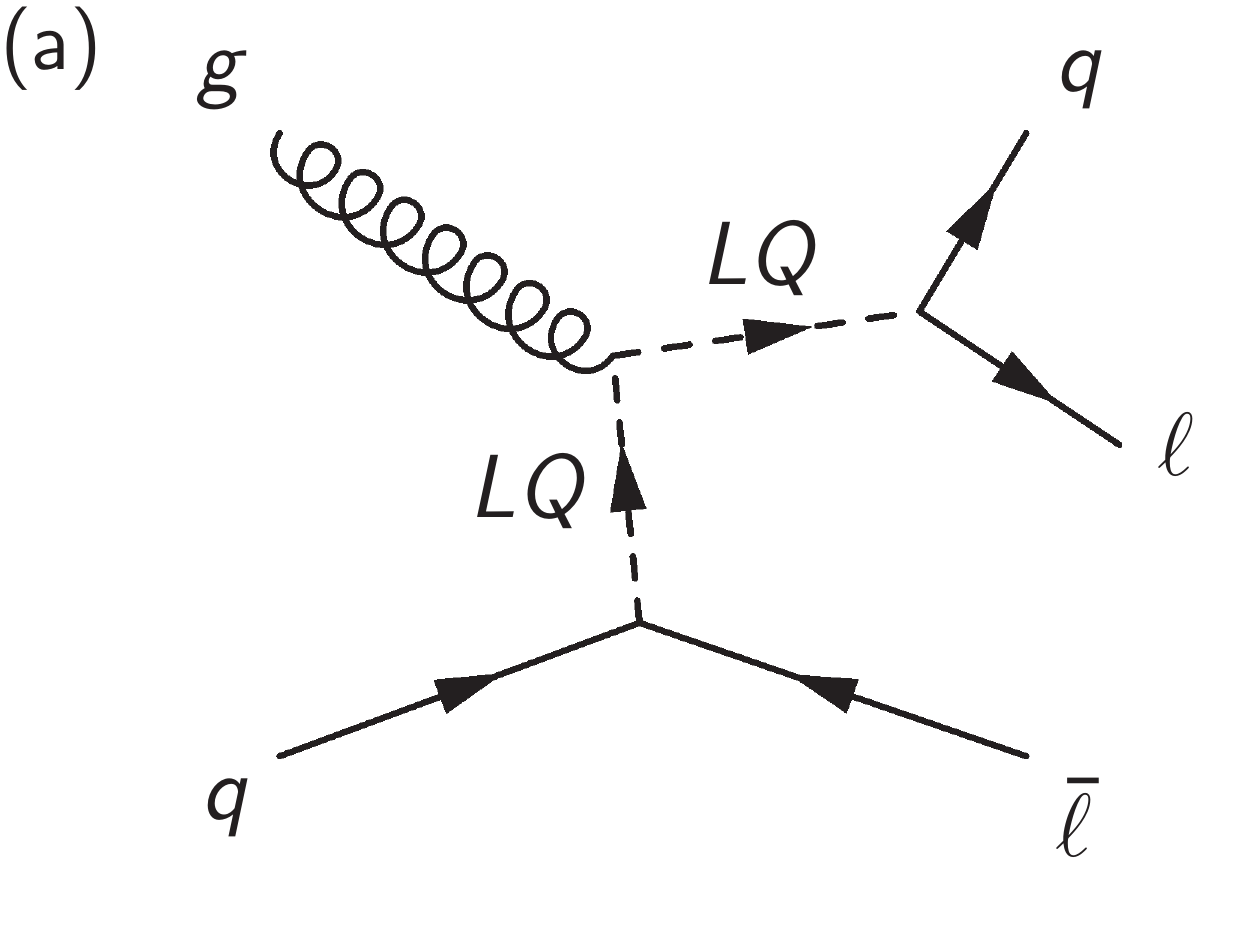}
       \includegraphics[scale=.45]{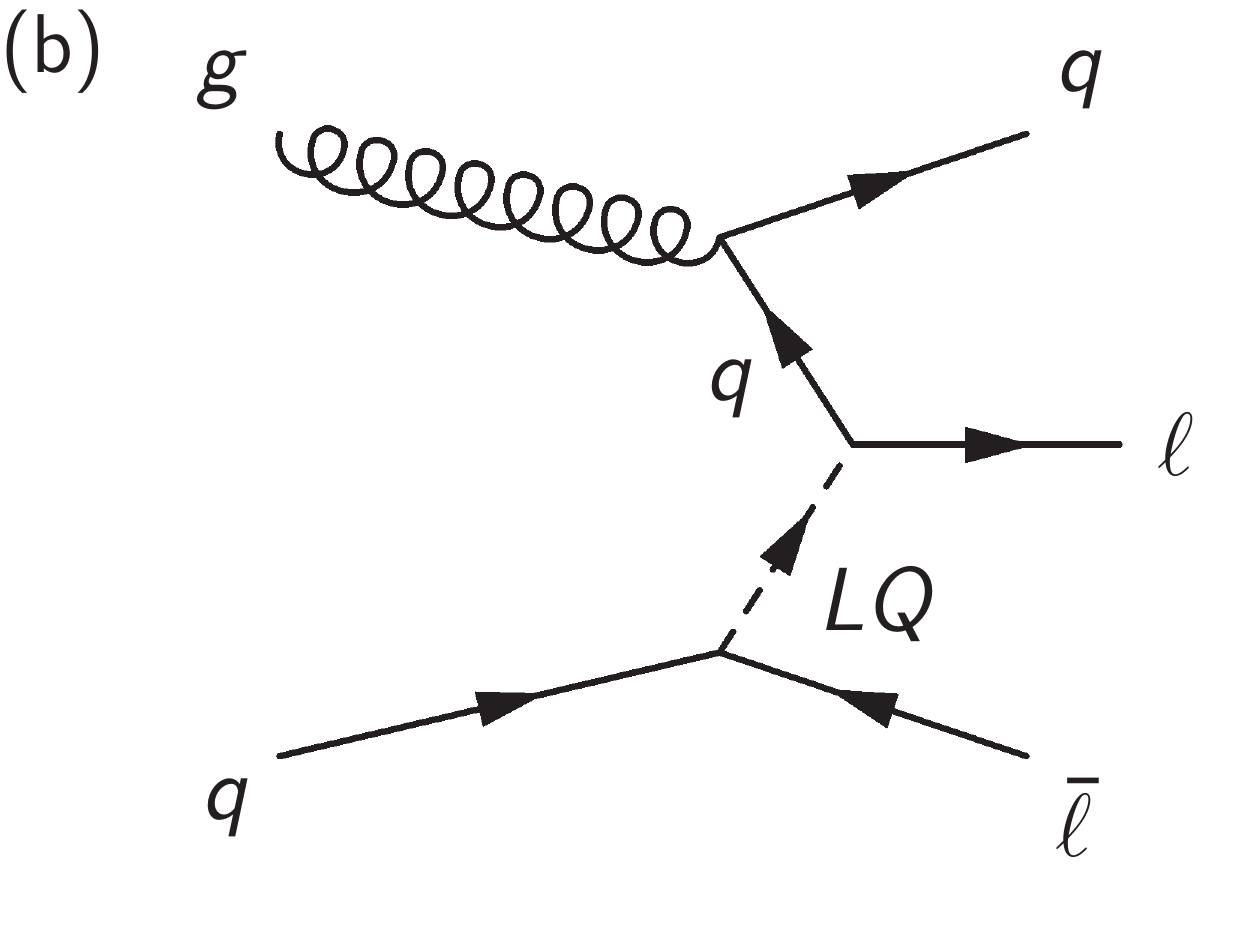}
       \caption{The $t$-channel LQ production diagrams with non-resonant components.  The diagram in (b) is completely non-resonant.
	 \label{figapp:NonResDiagram}}
\end{figure}

The main single leptoquark production mode at the LHC is the resonant diagram shown in Fig.~\ref{figapp:ResDiagram}.  However, significant contributions are made by the diagrams with non-resonant components shown in Fig.~\ref{figapp:NonResDiagram}.  These contributions increase with both the LQ mass and coupling; the invariant mass distribution of a first generation LQ, of mass $M_{\text{LQ}} = 1$ \TeV and coupling $\lambda=1.0$, possesses a tail extending to very low masses that is comparable to the peak in magnitude.  The reconstructed shape of the resonance peak itself is not strongly affected by $\lambda$.

Also, interference with the $\PQq \Pg \rightarrow \PQq \cPZ/\gamma^{*}\rightarrow \PQq \ell^{+} \ell^{-}$ SM process can occur at dilepton masses in the vicinity of the $\cPZ$ boson mass peak and at lower energies.  Treatments for this interference region and the above-described low-mass off-shell tail of the lepton-jet mass distribution are detailed in Section~\ref{selection}.

The final-state event signatures from the decays of singly produced LQs can be classified as either that of two charged leptons and a jet, where the LQ decays to a charged lepton and a quark, or of a charged lepton, missing transverse energy, and a jet, where the LQ decays into a neutrino and a quark.  The two signatures have branching fractions of $\beta$ and $1-\beta$, respectively.  For this study, and for $S^{R}_{0}$ type LQs, $\beta$ is 1.0, disregarding LQ decays to a neutrino and a quark.  Because the parton distribution functions (PDF) of the proton are dominated by the u and d quarks, the single production of LQs of second and third generation is suppressed.

The charged leptons can be electrons, muons, or taus, corresponding to the three generations of LQs.  In this paper two distinct signatures with charged leptons in the final state are considered: one with two high transverse momentum ($\PT$) electrons and one high-$\PT$ jet (denoted as $\Pe \Pe j$), and the other with two high--$\PT$ muons and one high-$\PT$ jet (denoted as $\Pgm \Pgm j$).

\section{The CMS detector}
\label{cms}

The central feature of the CMS apparatus is a superconducting solenoid of 6\unit{m} internal diameter, providing a magnetic field of 3.8\unit{T}.  Within the solenoid volume are a silicon pixel and strip tracker, a lead tungstate crystal electromagnetic calorimeter (ECAL), and a brass and scintillator hadron calorimeter, each composed of a barrel and two endcap sections.  Extensive forward calorimetry complements the coverage provided by the barrel and endcap detectors.  Muons are measured in gas-ionization detectors embedded in the steel flux-return yoke outside the solenoid.

The ECAL energy resolution for electrons with $\ET \approx 45$\GeV from $\Z \rightarrow \Pe \Pe$ decays is better than 2\% in the central pseudorapidity region of the ECAL barrel $(\abs{\eta} < 0.8)$, and is between 2\% and 5\% elsewhere. For low-bremsstrahlung electrons, where 94\% or more of their energy is contained within a $3{\times}3$ array of crystals, the energy resolution improves to 1.5\% for $\abs{\eta} < 0.8$~\cite{Chatrchyan:2013dga}.

Muons are measured in the pseudorapidity range $\abs{\eta}< 2.4$ with detection planes made using three technologies: drift tubes, cathode strip chambers, and resistive-plate chambers. Matching muon tracks derived from these measurements to tracks measured in the silicon tracker results in a relative \PT resolution for muons with $20 <\pt < 100\GeV$ of 1.3--2.0\% in the barrel and better than 6\% in the endcaps; the \pt resolution in the barrel is better than 10\% for muons with \pt up to 1\TeV~\cite{Chatrchyan:2012xi}.

The first level of the CMS trigger system, composed of custom hardware processors, uses information from the calorimeters and muon detectors to select the most interesting events. The high-level trigger (HLT) processor farm further decreases the event rate from around 100\unit{kHz} to around 400\unit{Hz}, before data storage.

The particle-flow event algorithm reconstructs and identifies each individual particle with an optimized combination of information from the various elements of the CMS detector. The energy of photons is directly obtained from the ECAL measurement, corrected for zero-suppression effects. The energy of electrons is determined from a combination of the electron momentum at the primary interaction vertex as determined by the tracker, the energy of the corresponding ECAL cluster, and the energy sum of all bremsstrahlung photons spatially compatible with originating from the electron track. The energy of muons is obtained from the curvature of the corresponding track. The energy of charged hadrons is determined from a combination of their momentum measured in the tracker and the matching ECAL and HCAL energy deposits, corrected for zero-suppression effects and for the response function of the calorimeters to hadronic showers. Finally, the energy of neutral hadrons is obtained from the corresponding corrected ECAL and HCAL energy.

A more detailed description of the CMS detector, together with a definition of the coordinate system used and the relevant kinematic variables, can be found in~\cite{Chatrchyan:2008zzk}.

\section{Data and simulation samples}
\label{samples}

The data were collected during the 8\TeV pp run in 2012 at the CERN LHC and correspond to an integrated luminosity of 19.6\fbinv.  In the $\Pe \Pe j$ channel, events are selected using a trigger that requires two electrons with $\PT > 33$\GeV and $|\eta| < 2.4$ and in the $\Pgm \Pgm j$ channel, events are selected using a trigger that requires one muon with $\PT > 40$\GeV and $|\eta| < 2.1$.

Simulated samples for the signal processes are generated for a range of leptoquark mass hypotheses between 300 and 3300\GeV and coupling hypotheses between 0.4 and 1.0 in the $\Pe \Pe j$ channel, and a range of leptoquark mass hypotheses between 300 and 1800\GeV and a coupling hypothesis of 1.0 in the $\Pgm \Pgm j$ channel.  Production of LQs in the $\Pgm \Pgm j$ channel is suppressed because of the proton PDF as discussed in Section~\ref{introduction}.

The main sources of background are $\ttbar$, $\cPZ/\gamma^* + {\rm jets}$, $\PW + {\rm jets}$, diboson $(\cPZ\cPZ , \cPZ\PW, \PW\PW) + {\rm jets}$, single top quark, and QCD multijet production.  The $\ttbar + {\rm jets}$ background shape is estimated from a study based on data described in Section~\ref{backgrounds}; the simulation sample for the normalization of the $\ttbar + {\rm jets}$ background as well as the samples for the $\cPZ/\gamma^* + {\rm jets}$ and $\PW + {\rm jets}$ backgrounds are generated with \MADGRAPH 5.1~\cite{MadGraph}.  Single top quark samples ($\PQs$-, $\PQt$-channels, and $\PW$ boson associated production) are generated with \POWHEG 1.0~\cite{POWHEG,Alioli:2009je,Frixione:2007vw,Nason:2004rx} and diboson samples are generated with \PYTHIA (version 6.422)~\cite{PYTHIA} using the Z2 tune~\cite{tunez2}.  The QCD multijet background is estimated from data.

\begin{figure}[!h]
       \centering
       {\includegraphics[width=.45\textwidth]{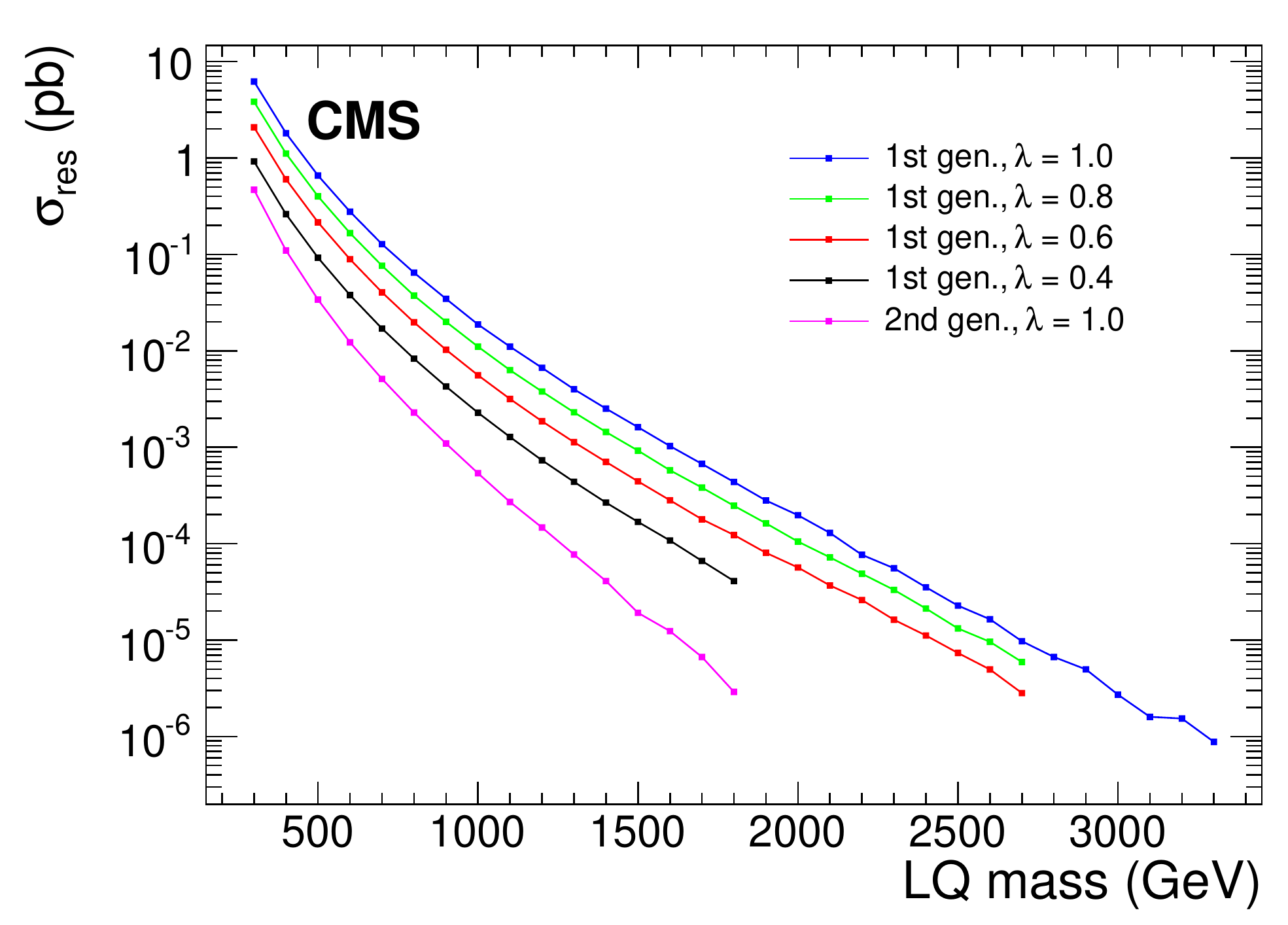}}
       \caption{Cross sections for single LQ production, calculated at LO in {{\textsc{CalcHEP}}\xspace} and scaled by the acceptance of the requirements described in Section~\ref{selection}, as a function of the LQ mass in \GeVns.
	 \label{figapp:CrossSections}}
\end{figure}

For the simulation of signal samples, the {{\textsc{CalcHEP}}\xspace}~\cite{hepmdb} generator is used for calculation of the matrix elements. The signal cross sections are computed at leading order (LO) with {{\textsc{CalcHEP}}\xspace} and are listed in Table~\ref{tab:sigxsectionscombined} in the appendix.  Blank entries were not considered because of the small size of the cross section. The resonant cross sections $\sigma_{\text{res}}$ are shown in Fig.~\ref{figapp:CrossSections} and are defined by the kinematics selections given in Section~\ref{selection}.

The \PYTHIA and \MADGRAPH simulations use the CTEQ6L1~\cite{CTEQ} PDF sets, those produced with {{\textsc{CalcHEP}}\xspace} use the CTEQ6L PDFs, and the \POWHEG simulation uses the CTEQ6m set.  All of the simulations use \PYTHIA for the treatment of parton showering, hadronization, and underlying event effects.  For both signal and background simulated samples, the simulation of the CMS detector is based on the \GEANTfour package~\cite{GEANT4}.  All simulated samples include the effects of extra collisions in a single bunch crossing as well as collisions from nearby bunch crossings (out-of-time pileup and in-time pileup, respectively).  The pileup profiles in simulation are reweighted to the distributions of the reconstructed vertices per bunch crossing in data collected by the CMS detector~\cite{CMS-PAS-JME-12-002}.

In the $\Pe \Pe j$ channel, the background and signal are rescaled by a uniform trigger efficiency scale factor of 0.996, which is measured in~\cite{CMS-PAPERS-EXO-12-061}.
In the $\Pgm \Pgm j$ channel, the background and signal are rescaled by muon $\eta$-dependent efficiency factors of 0.94 $(|\eta| \le 0.9$), 0.84 ($0.9 < |\eta| \le 1.2$), and 0.82 ($1.2 < |\eta| \le 2.1$).  An uncertainty of 1\% is assigned to these factors to account for variations during data-taking periods and statistical uncertainties.

\section{Event reconstruction}
\label{objectreco}

Muons are reconstructed as tracks in the muon system that are ``globally'' matched to reconstructed tracks in the tracking system~\cite{Chatrchyan:2012xi}.  Muons are required to have $\PT > 45$\GeV and $|\eta| < 2.1$.  Additionally, they are required to satisfy a set of criteria that is optimized for high $\PT$; they are reconstructed as ``global'' muons with tracks associated to hits from at least two muon detector planes together with at least one muon chamber hit that is included in the "global" track fit~\cite{Chatrchyan:2012xi}.  To perform a precise measurement of the \PT and to reduce background from muons from secondary decays in flight, at least eight hits are required in the tracker and at least one in the pixel detector.   To minimize background from muons from cosmic ray backgrounds, the transverse impact parameter with respect to the primary vertex is required to be less than 2 mm and the longitudinal distance is less than 5 mm.  Muons are required to be isolated by applying an upper threshold on the relative tracker isolation of 0.1.  The relative tracker isolation is defined as the ratio of the \PT of all tracks in the tracker coming from the same vertex, excluding the muon candidate track, in a cone of {$\Delta R = \sqrt{\smash[b]{(\Delta \phi) ^2 + (\Delta \eta) ^2}} = 0.3$} (where $\phi$ is the azimuthal angle in radians) around the muon candidate track, and the muon \PT.

Electrons are required to have a reconstructed track in the central tracking system that is matched in $\eta$ and $\phi$ to a cluster of ECAL crystals that has a shape consistent with an electromagnetic shower.  The transverse impact parameter of the track with respect to the primary vertex is required to be less than 2 mm for electrons in the barrel ($\abs{\eta} < 1.442$) and less than 5 mm for electrons in the endcap ($\abs{\eta} > 1.560$).  Electrons are required to be isolated from reconstructed tracks other than the matched track in the central tracking system and from additional energy deposits in the calorimeter.  The transverse momentum sum of all tracks in a cone of $\Delta R = 0.5$ around the electron candidate's track and coming from the same vertex must be less than 5\GeV.  Also, the transverse energy sum of the calorimeter energy deposits falling in the cone of $\Delta R = 0.5$ must be less than 3\% of the candidate's transverse energy.  An additional contribution accounting for the average contribution of other proton-proton collisions in the same bunch crossing is added to this sum.  To reject electrons coming from photon conversions within the tracker material, the reconstructed electron track is required to have hits in all pixel layers.  Electrons in the analysis have $\PT > 45$\GeV and $|\eta| < 2.1$ to match the muon requirements (excluding the transition region between barrel and endcap detectors, $1.442  < \abs{\eta} < 1.560 $).  Selection criteria for electron identification and isolation optimized for high energies are also applied~\cite{CMS-PAPERS-EXO-12-061}.

Jets are reconstructed with the CMS particle-flow algorithm~\cite{CMS-PAS-PFT-09-001,CMS-PAS-PFT-10-001}, which measures
stable particles by combining information from all CMS subdetectors. The jet reconstruction algorithm used
in this paper is the anti-k$\rm _T$~\cite{Cacciari:2008gp,fastjetmanual} algorithm with a
distance parameter $0.5$, which only considers tracks associated to the primary vertex. Jet momentum is determined as the vectorial sum of all particle momenta in the jet, and is found from simulation to be within 5\% to 10\% of the true momentum over the whole \pt spectrum and detector acceptance.  An offset correction is applied to jet energies to take into account the contribution from additional proton-proton interactions within the same bunch crossing. Jet energy corrections are derived from simulation, and are confirmed with in situ measurements of the energy balance in dijet and photon+jet events~\cite{CMS-PAPERS-JME-10-011}. Additional selection criteria are applied to each event to remove spurious jet-like features originating from isolated noise patterns in certain HCAL regions.  The jet energy resolution amounts typically to 15\% at 10\GeV, 8\% at 100\GeV, and 4\% at 1\TeV, to be compared to about 40\%, 12\%, and 5\% obtained when the calorimeters alone are used for jet clustering~\cite{CMS-PAS-PFT-09-001}.

Jets are required to have $\PT>45$\GeV, $|\eta| < 2.4$, and an angular separation from leptons of $\Delta R > 0.3$.

\section{Event selection}
\label{selection}

We require that events in both the $\Pe \Pe j$ and $\Pgm \Pgm j$ channels contain at least two leptons and at least one jet that satisfy the above identification criteria.  Additional kinematic requirements are applied to remove regions in which the trigger and identification criteria are not at plateau efficiency and to reduce large backgrounds.  This creates a basic preselection region:  the jet $\PT$ must be larger than 125\GeV, the dilepton invariant mass $M_{\ell\ell} $ must be larger than 110\GeV, and the scalar sum of transverse momenta of objects in the event ($\ST = \PT(\ell_1)+\PT(\ell_2)+\PT(j_1)$) is required to exceed 250\GeV, where $\ell_1$ is the highest $\PT$ lepton in the event, $\ell_2$ is the second-highest $\PT$ lepton, and $j_1$ is the highest $\PT$ jet.  The two leptons in the events are required to have opposite charges.

After this initial selection, a final selection is optimized for each channel separately by maximizing $S/\sqrt{S+B}$, where $S$ is the number of signal events in the simulation passing a given selection and $B$ is the number of background events in the simulation passing the same selection.  We optimize for each LQ mass hypothesis by varying the requirements on $M_{\ell j}$ and $\ST$.  Here $M_{\ell j}$ is defined as the higher of the two possible lepton-jet mass combinations.

As discussed in Section~\ref{introduction}, owing to the unique aspects of single LQ decays, two generator level requirements are applied to the simulated signal samples.  The first is $M_{\ell\ell} > 110$\GeV, to remove LQ decays that are in the Z boson interference region.  The second is a requirement on $M_{\ell j}$, chosen to remove the $t$-channel diagram contributions in the low-mass off-shell region, while preserving most of the resonant signal.  This requirement is set at $M_{\ell j} > 0.67 \, M_{\text{LQ}}$ for the first-generation studies and $M_{\ell j} > 0.75 \, M_{\text{LQ}}$ for the second-generation studies.  The thresholds for $M_{\ell j}$ were chosen separately for each channel, because of the differences in the distribution shape.  The dilepton invariant mass requirement at the generator level precisely matches the reconstruction level requirement at the preselection.  These two requirements define the resonant region.  Cross sections at the generator level before and after these requirements are provided in Table~\ref{tab:sigxsectionscombined}, in the appendix.

The $\Pe \Pe j$ channel selection after optimization is identical for all couplings.  The threshold on $\ST$ starts at 250\GeV for $M_{\text{LQ}}=300$\GeV and increases linearly until it reaches a plateau value of 900\GeV at $M_{\text{LQ}}=1125$\GeV.  The $M_{\ell j}$ threshold starts at 200\GeV for $M_{\text{LQ}}=300$\GeV and increases linearly until it plateaus at 1900\GeV above $M_{\text{LQ}}=2000$\GeV.  In the $\Pgm \Pgm j$ channel after optimization the threshold on $\ST$ starts at 300\GeV for $M_{\text{LQ}}=300$\GeV and increases linearly until it plateaus at 1000\GeV above $M_{\text{LQ}}=1000$\GeV.  The $M_{\ell j}$ threshold starts at 200\GeV for $M_{\text{LQ}}=300$\GeV and increases linearly until it plateaus at 800\GeV above $M_{\text{LQ}}=900$\GeV.  The exact threshold values are listed in Tables \ref{tab:optimizedcuts1p0} and \ref{tab:optimizedcutscmu} in the appendix.

\section{Background estimations}
\label{backgrounds}

The SM processes that mimic the signal signature are $\cPZ/\gamma^* + {\rm jets}$, $\ttbar$, single top quark, diboson $ + {\rm jets}$, $\PW + {\rm jets}$, and QCD multijets events where the jets are misidentified as leptons.  The dominant contributions come from the former two processes, whereas the other processes provide minor contributions to the total number of background events.

The contribution from the $\cPZ/\gamma^* + {\rm jets}$ background is estimated with a simulated sample that is normalized to agree with data at preselection in the $\cPZ$-enriched region of $80 < M_{\ell\ell} < 100$\GeV, where $M_{\ell\ell}$ is the dilepton invariant mass.  With this selection the data sample (with non-$\cPZ/\gamma^* + {\rm jets}$ simulated samples subtracted) is compared to $\cPZ/\gamma^* + {\rm jets}$ in simulation.  The resulting scale factor, representing the ratio of the measured yield to the predicted yield, is $R_{\cPZ} = 0.98 \pm 0.01 \stat$ in both the $\Pe \Pe j$ and $\Pgm \Pgm j$ channels.  This scale factor is then applied to the simulated $\cPZ/\gamma^* + {\rm jets}$ sample in the signal region of $M_{\ell\ell} > 110$\GeV.  In order to account for possible mismodeling of the $\PT(\ell\ell)$ spectrum of the $\cPZ/\gamma^* + {\rm jets}$ background sample, where $\PT(\ell\ell)$ is the scalar sum of the two highest \PT leptons in the event, we perform a bin-by-bin rescaling of yields at preselection and full selection by scale factors measured in an inverted $M_{\ell\ell}$ selection ($M_{\ell\ell} < 110$\GeV).  These scale factors differ from unity by 1\% to 10\%, depending on the $\PT(\ell\ell)$ bin, and are applied to the $\cPZ/\gamma^* + {\rm jets}$ sample in the signal region of $M_{\ell\ell} > 110$\GeV.

We estimate the $\ttbar$ background with a $\ttbar$-enriched $\Pe \Pgm$ sample in data, selected using the single muon trigger.  We use a selection that is identical to our signal selection in terms of kinematics requirements, except that we require at least a single muon and a single electron rather than requiring two same-flavor leptons.   The $\Pe \Pgm$ sample is considered to be signal-free, because limits on flavor changing neutral currents imply that LQ processes do not present a different-flavor decay topology~\cite{Buchmuller:1986iq,FCNC}. The $\ttbar$ background is largely dominant in the $\Pe \Pgm$ sample with respect to the other backgrounds.
This background is expected to produce the $\Pe \Pe$ ($\Pgm \Pgm$) final state with half the probability of the $\Pe \Pgm$ final state, thus the $\Pe \Pgm$ sample is scaled by a factor of $1/2$.
This factor is multiplied by the ratio of electron (muon) identification and isolation efficiencies, $R_{\Pe \Pe /\Pe \Pgm}$ ($R_{\Pgm \Pgm /\Pe \Pgm}$).
The estimate is further scaled by the ratio of the double-electron trigger efficiency and the single muon efficiency, $R_{\text{trig},\Pe \Pe}$ in Eq.~\ref{eq:ttbarratio}, or by the ratio of the efficiency of the single muon trigger in dimuon final states and the single muon efficiency, $R_{\text{trig},\Pgm \Pgm}$ in Eq.~\ref{eq:ttbarratiomumu}.
The resulting estimates of the number of $\ttbar$ events in the $\Pe \Pe$ and $\Pgm \Pgm$ channels are

\begin{equation}
N^{\ttbar, \text{est}}_{\Pe \Pe} = (N_{\Pe \Pgm}^{\text{data}}-N_{\Pe \Pgm}^{\text{non-}\ttbar \text{ sim}}) \, \frac{1}{2}\, R_{\Pe \Pe/\Pe \Pgm} \, R_{\text{trig,} \Pe \Pe}
,
\label{eq:ttbaree}
\end{equation}

\begin{equation}
N^{\ttbar, \text{est}}_{\Pgm \Pgm} = (N_{\Pe \Pgm}^{\text{data}}-N_{\Pe \Pgm}^{\text{non-}\ttbar \text{ sim}}) \, \frac{1}{2}\, R_{\Pgm \Pgm/\Pe \Pgm} \, R_{\text{trig,} \Pgm \Pgm}
,
\label{eq:ttbarmumu}
\end{equation}
with
\begin{equation}
R_{\text{trig},\Pe \Pe} = \frac{\epsilon_{\Pe \Pe}}{\epsilon_{\Pgm}}
,
\label{eq:ttbarratio}
\end{equation}
\begin{equation}
R_{\text{trig},\Pgm \Pgm } = \frac{1-(1-\epsilon_{\Pgm})^{2}}{\epsilon_{\Pgm}} = 2-\epsilon_{\Pgm}
,
\label{eq:ttbarratiomumu}
\end{equation}

where $\epsilon_{\Pgm}$ and $\epsilon_{\Pe \Pe}$ are the single-muon trigger and double-electron trigger efficiencies, respectively, and $N_{\Pe \Pgm}^{\text{data}}$ and $N_{\Pe \Pgm}^{\text{non-}\ttbar \text{ sim}}$ are the numbers of $\Pe \Pgm$ events observed in data and estimated from backgrounds other than $\ttbar$, respectively.  $R_{\text{trig},\Pgm \Pgm }$ is the ratio of the efficiency of a single muon trigger on a dimuon sample over the efficiency on a single muon sample (the numerator is the likelihood of failure on two muons).

The contribution from QCD multijet processes is determined by a method that makes use of the fact that neither signal events nor events from other backgrounds produce final states with same-charge leptons at a significant level.  We create four selections, with both opposite-sign (OS) and same-sign (SS) charge requirements, as well as isolated and non-isolated requirements.  Electrons in isolated events must pass the isolation criteria optimized for high-energy electrons~\cite{CMS-PAPERS-EXO-12-061} and muons are required to have a relative tracker isolation less than 0.1, as discussed in Section~\ref{objectreco}.  Non-isolated events are those with leptons failing these criteria.  The four selections are as follows,
\begin{equation}
\sloppy
\left( \begin{array}{cc}
  A & B  \\
  C & D
\end{array} \right) =
\left( \begin{array}{cc}
  \text{OS+isolated} & \text{OS+non-isolated}  \\
  \text{SS+isolated} & \text{SS+non-isolated}
\end{array} \right).
\label{eq:ACBDMatrix}
\end{equation}

The shape of the background is taken from the SS region with isolation requirements, and the normalization is obtained from the ratio between the number of OS events and the number of SS events in the non-isolated selection.  Thus, the number of events, $N^{\text{QCD, est}}$, is estimated by
\begin{equation}
\small{
N^{\text{QCD, est}} = r_{B/D} \, N_{C}^{(\text{data -- non-QCD sim})}
},
\label{eq:QCD_ABCDeq}
\end{equation}
where $N_{C}^{(\text{data -- non-QCD sim})}$ is the number of events in region $C$ of Eq. (\ref{eq:ACBDMatrix}) and $r_{B/D}$ is the ratio of the number of events (measured in data with simulated non-QCD backgrounds subtracted) in regions $B$ and $D$.  The result is that QCD multijet processes account for 2\% (1\%) of the total SM background in the $\Pe \Pe j$ ($\Pgm \Pgm j$) channel.

The contributions of the remaining backgrounds (diboson+jets, \PW+jets, single top quark) are small and are determined entirely from simulation.

\begin{figure}[!htb]
       \centering
       {\includegraphics[width=.45\textwidth]{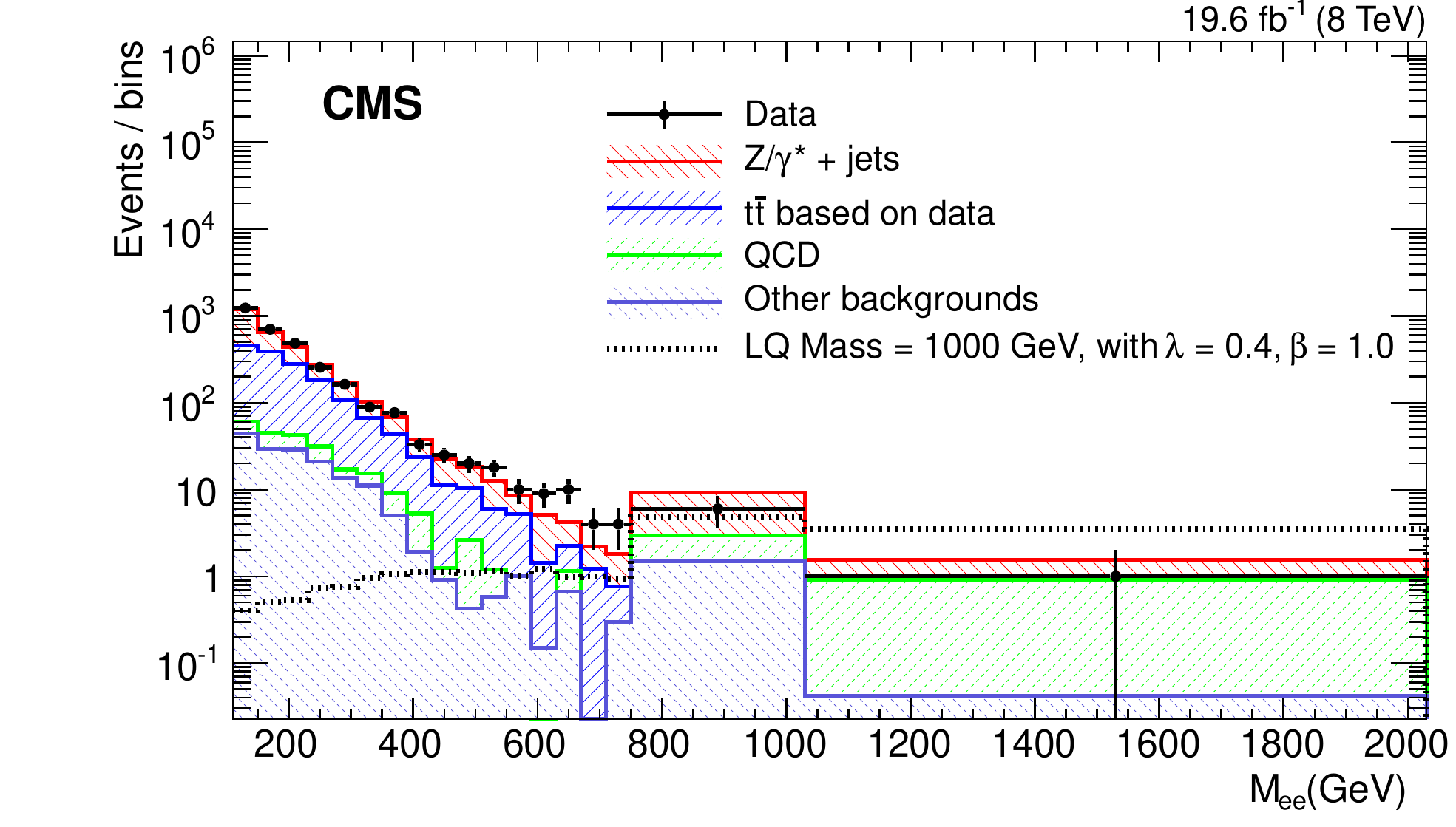}}
       {\includegraphics[width=.45\textwidth]{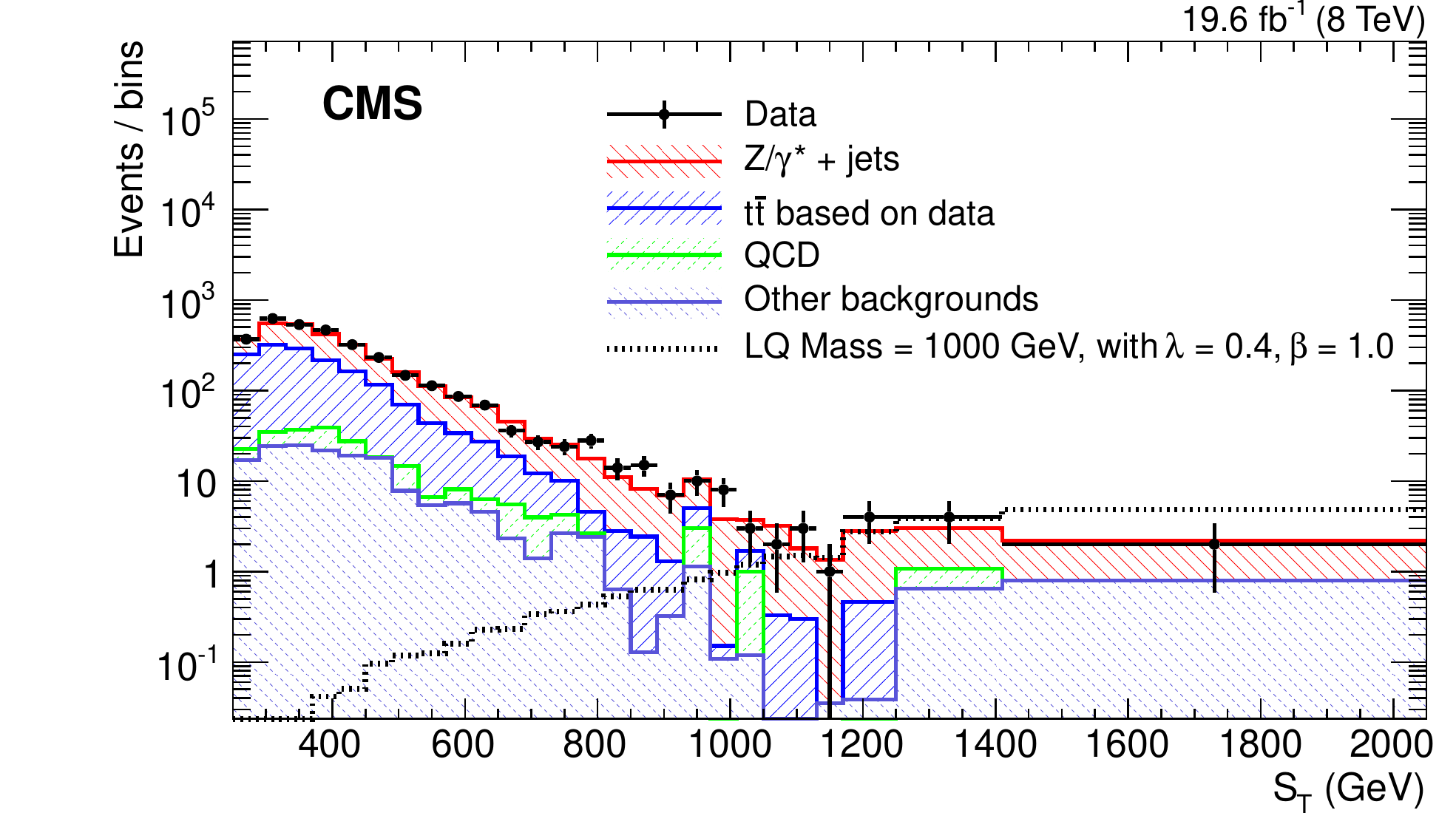}}
       {\includegraphics[width=.45\textwidth]{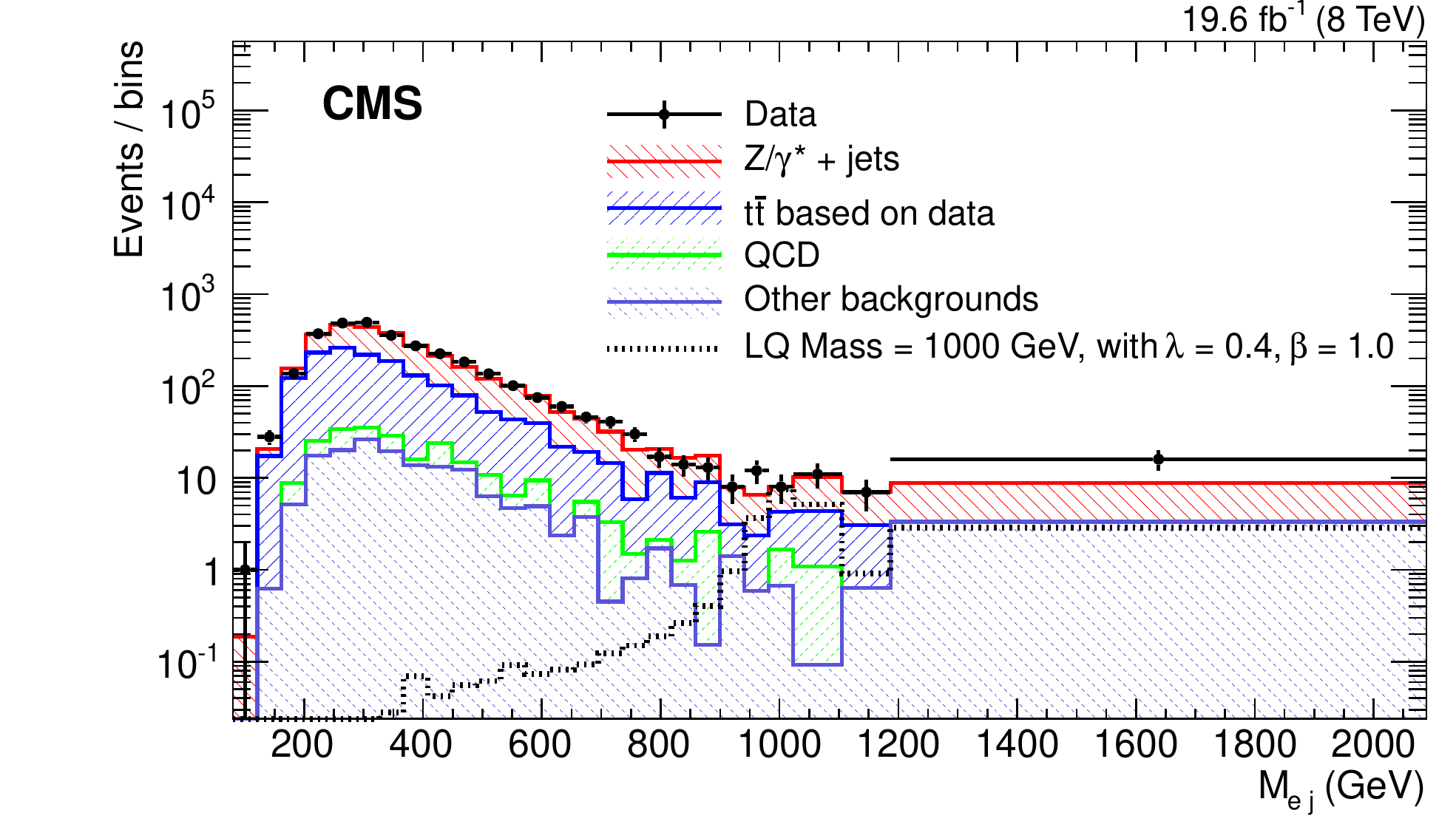}}
       \caption{Distributions of $M_{\Pe \Pe}$ (\cmsTopLeft), $\ST$ (\cmsTopRight), and $M_{\Pe j}$ (\cmsBottom) at preselection in the $\Pe \Pe j$ channel.  ``Other backgrounds'' include diboson, \PW $+$ jets, and single top quark contributions.  The points represent the data and the stacked histograms show the expected background contributions.  The open histogram shows the prediction for an LQ signal for $M_{\text{\text{LQ}}}$ = 1000\GeV and $\lambda=0.4$.  The horizontal error bars on the data points represent the bin width.  The last bin includes overflow. \label{figapp:elemisc}}
\end{figure}

\begin{figure}[!htb]
       \centering
       {\includegraphics[width=.45\textwidth]{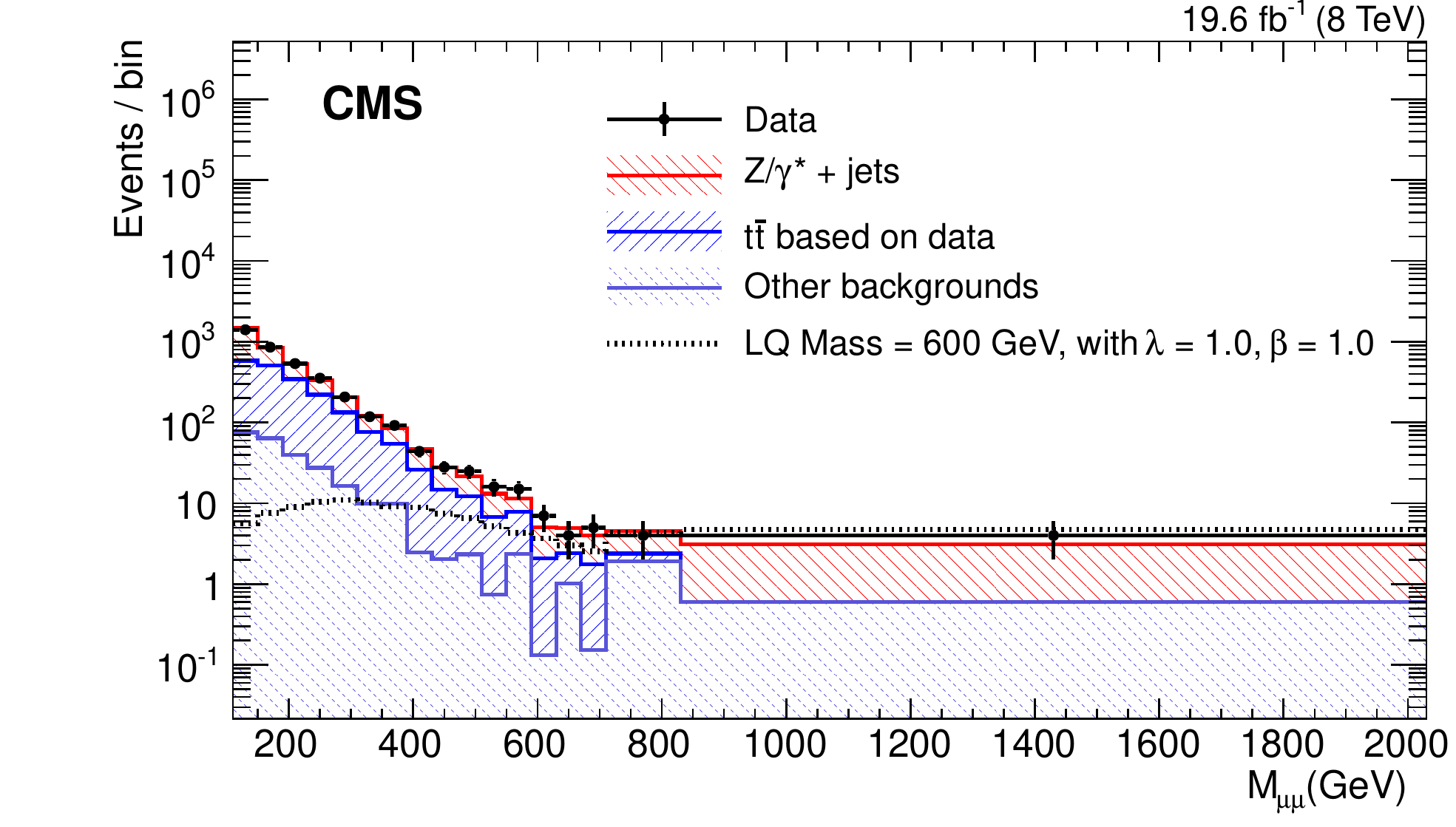}}
       {\includegraphics[width=.45\textwidth]{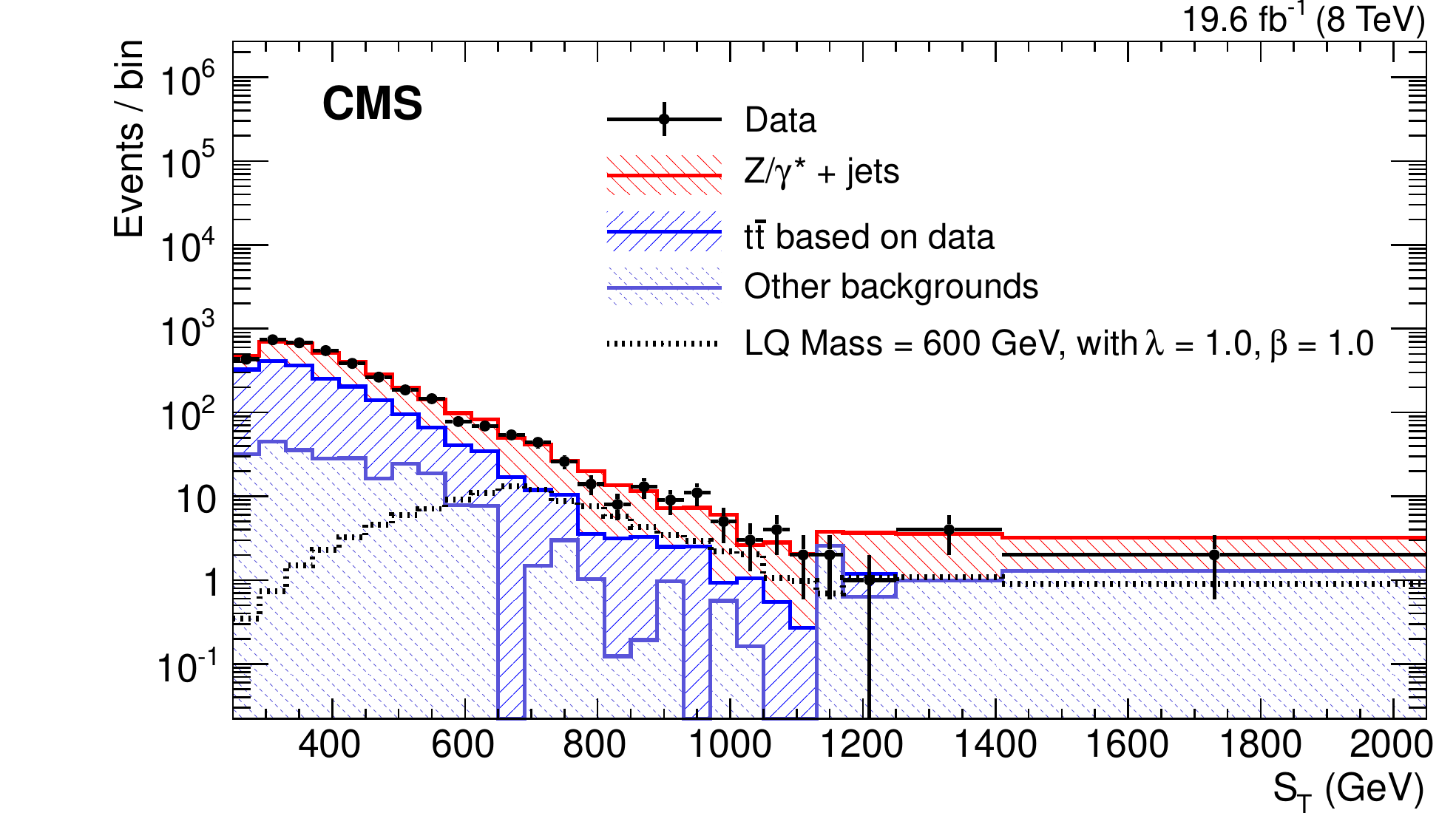}}
       {\includegraphics[width=.45\textwidth]{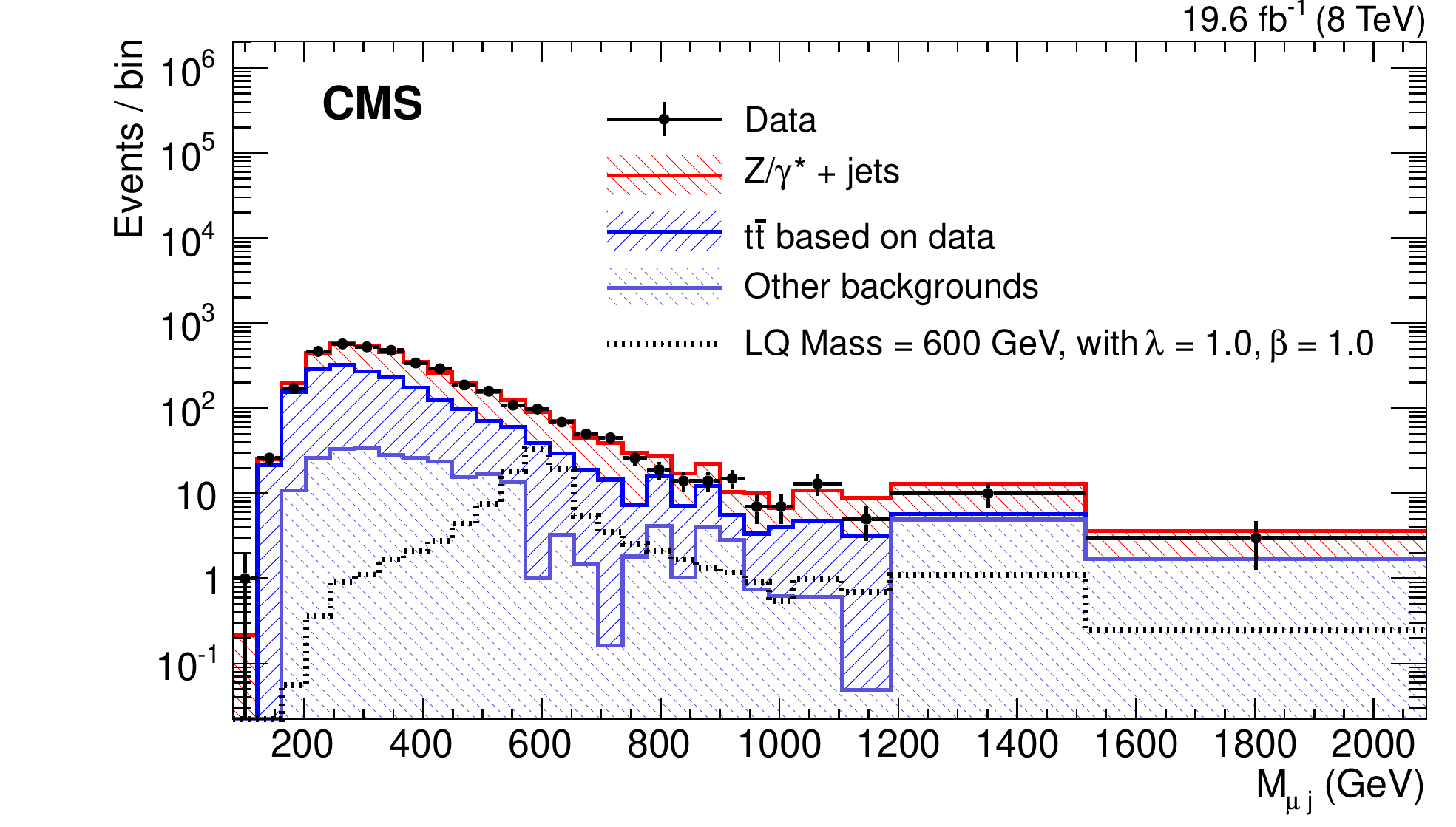}}
       \caption{Distributions of $M_{\Pgm \Pgm}$ (\cmsTopLeft), $\ST$ (\cmsTopRight), and $M_{\Pgm j}$ (\cmsBottom) at preselection in the $\Pgm \Pgm j$ channel.  ``Other backgrounds'' include diboson, \PW $+$ jets, single top quark, and QCD multijet contributions.  The points represent the data and the stacked histograms show the expected background contributions.  The open histogram shows the prediction for an LQ signal for $M_{\text{\text{LQ}}}$ = 600\GeV and $\lambda=1.0$.  The horizontal error bars on the data points represent the bin width.  The last bin includes overflow. \label{figapp:muonmisc}}
\end{figure}

The preselection level distributions in $M_{\ell \ell}$, $\ST$, and $M_{\ell j}$ are shown in Figs.~\ref{figapp:elemisc} and~\ref{figapp:muonmisc} for the observed data and estimated backgrounds, where they are compared with a signal LQ mass of 1000\GeV in the $\Pe \Pe j$ channel, and with a signal LQ of mass 600\GeV, in the $\Pgm \Pgm j$ channel.  In all plots the $\cPZ/\gamma^* + {\rm jets}$ prediction is normalized to data and the $\ttbar$ prediction is taken from the study based on data.  Data and background are found to be in agreement.  The numbers of events selected in data and in the backgrounds at each final selection (for each hypothesis mass) are shown in Tables~\ref{tab:eenumbers1p0},~\ref{tab:eesignumbers1p0}, and~\ref{tab:mumunumbers} in the appendix.

\begin{figure}[!htb]
       \centering
       {\includegraphics[width=.45\textwidth]{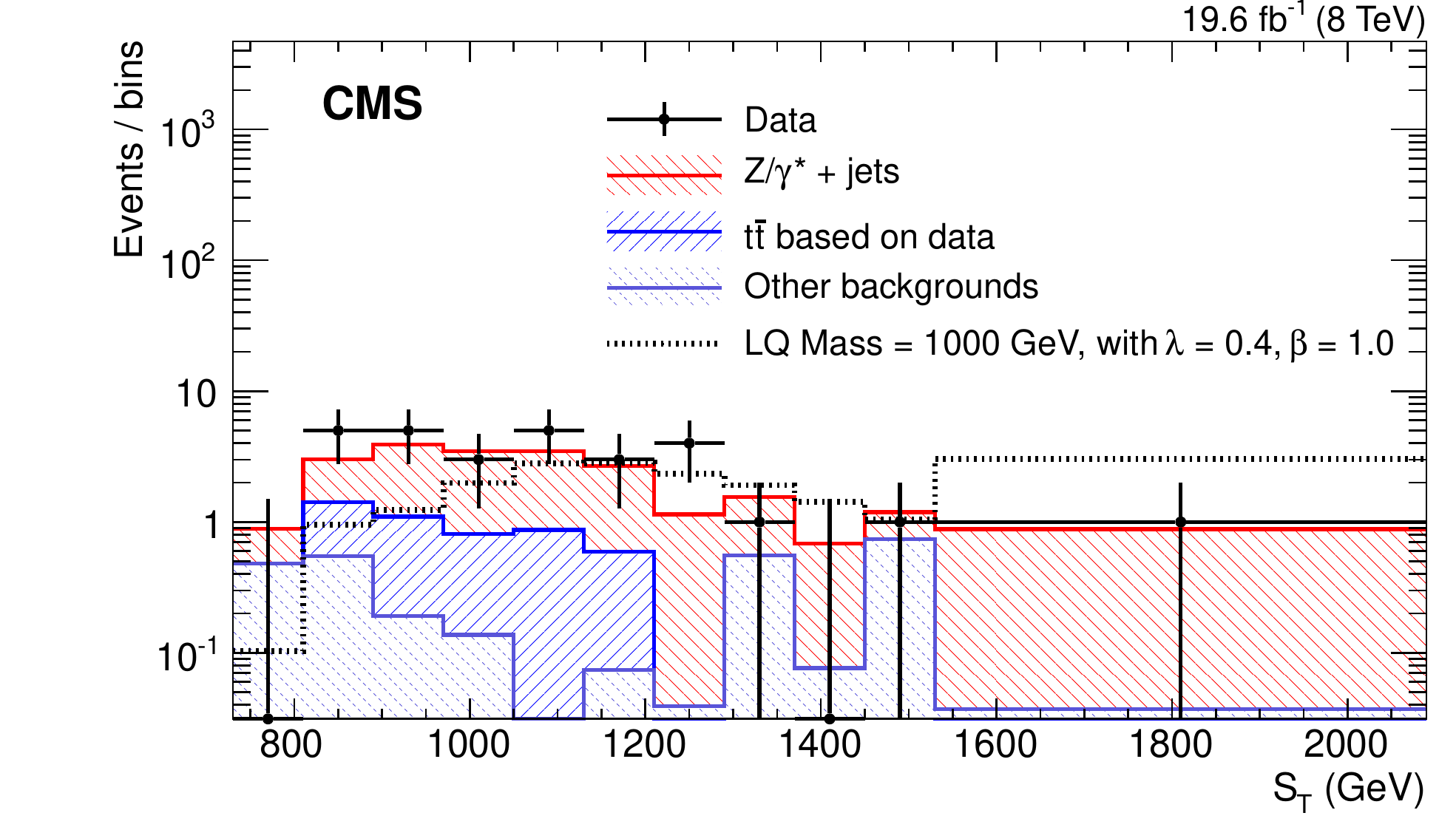}}
       {\includegraphics[width=.45\textwidth]{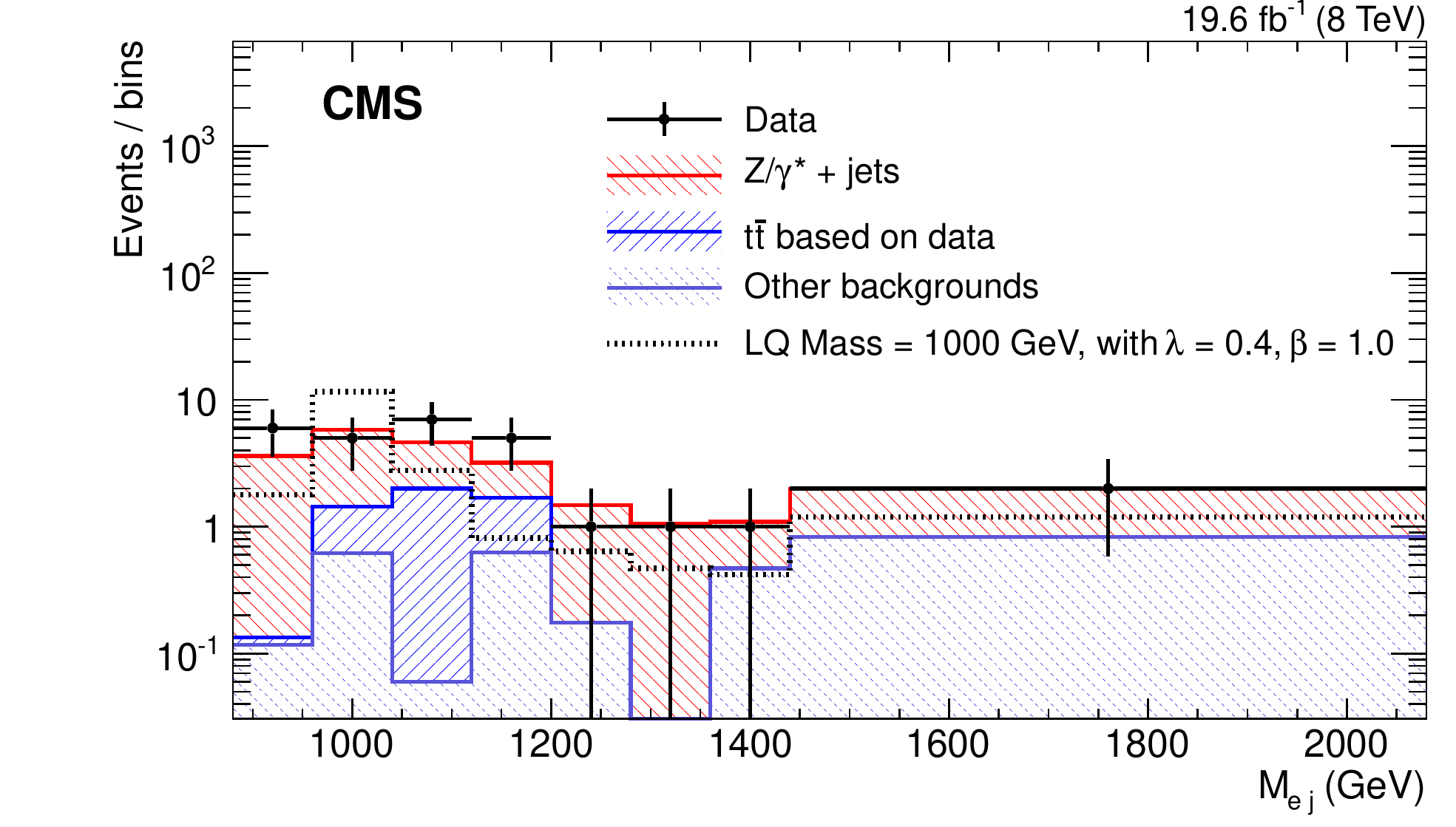}}
       \caption{Distributions of $\ST$ and $M_{\Pe j}$ at final selection, in the $\Pe \Pe j$ channel .  The points represent the data and the stacked histograms show the expected background contributions.  The open histogram shows the prediction for an LQ signal for $M_{\text{\text{LQ}}}=1000\GeV$ and $\lambda=0.4$.  The horizontal error bars on the data points represent the bin width.  The last bin includes overflow. \label{figapp:elefinalsel}}
\end{figure}

\begin{figure}[!htb]
       \centering
       {\includegraphics[width=.45\textwidth]{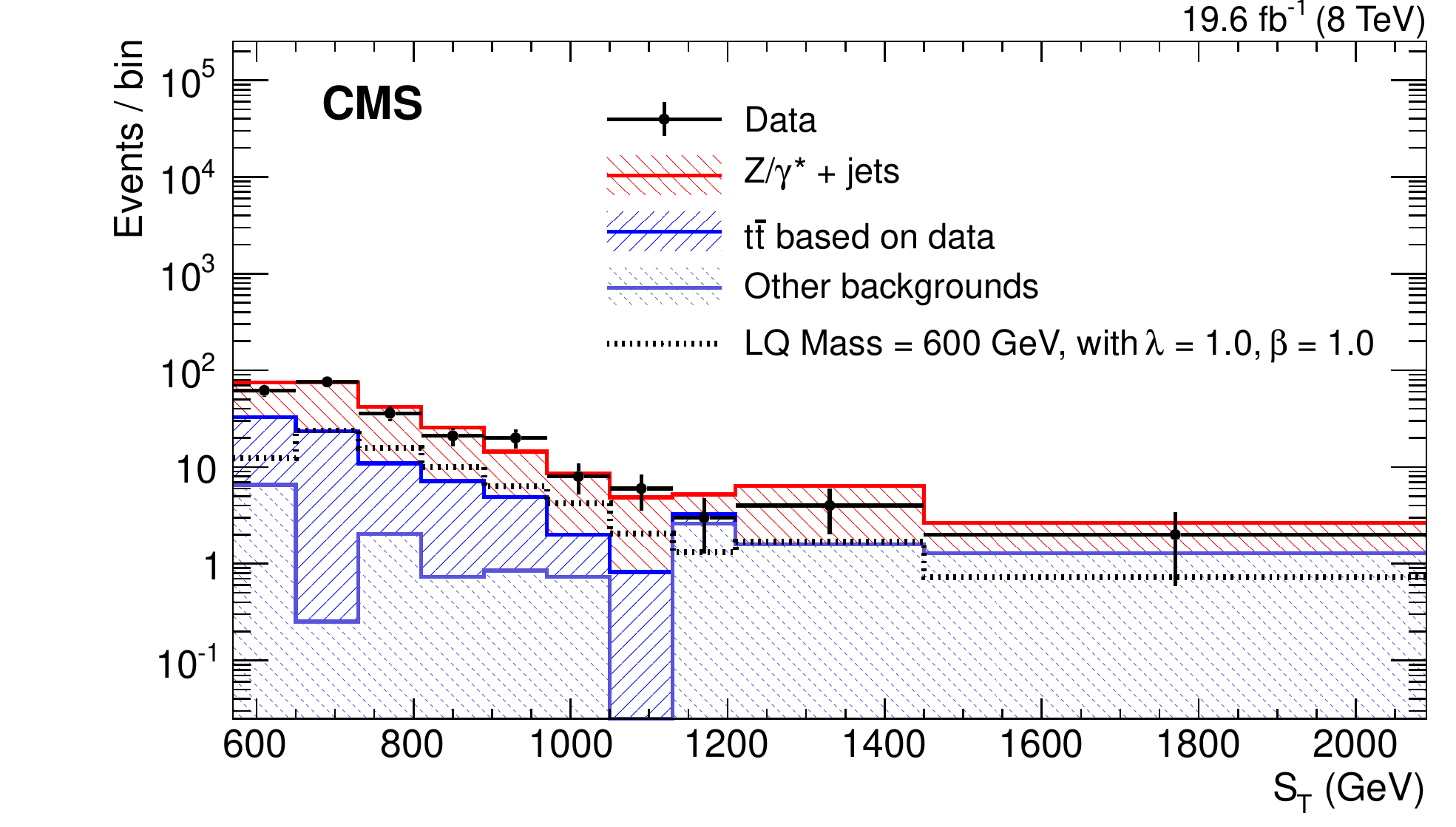}}
       {\includegraphics[width=.45\textwidth]{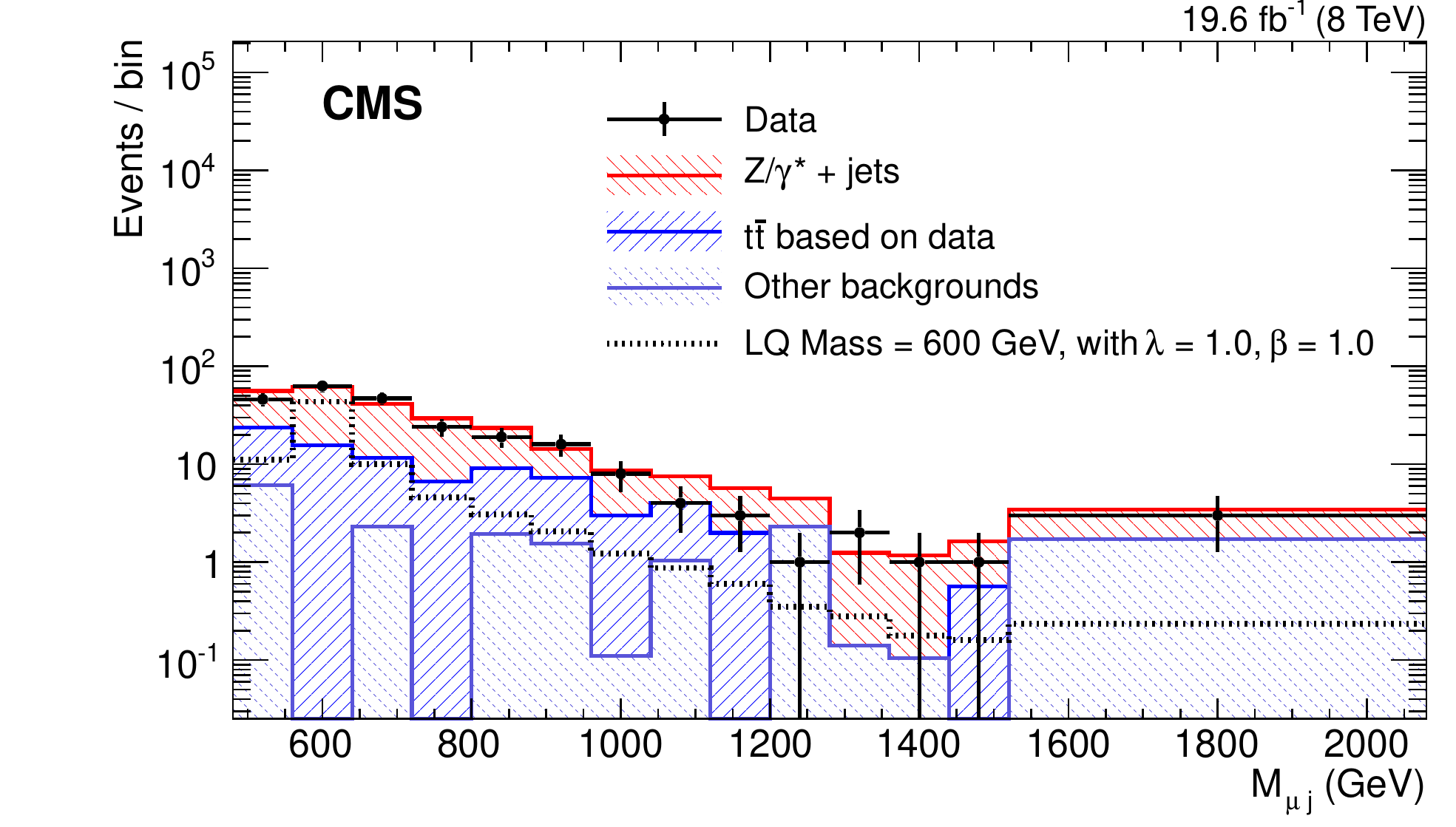}}
       \caption{Distributions of $\ST$ and $M_{\Pgm j}$ at final selection, in the $\Pgm \Pgm j$ channel.  The points represent the data and the stacked histograms show the expected background contributions.  The open histogram shows the prediction for an LQ signal for $M_{\text{\text{LQ}}}=600\GeV$ and $\lambda=1.0$.  The horizontal error bars on the data points represent the bin width.  The last bin includes overflow. \label{figapp:mufinalsel}}
\end{figure}

The observed data and background predictions are compared after final selection for $\lambda = 0.4$ and a signal LQ mass of 1000\GeV in the $\Pe \Pe j$ channel and a signal LQ mass of 600\GeV in the $\Pgm \Pgm j$ channel and are shown in Figs.~\ref{figapp:elefinalsel} and~\ref{figapp:mufinalsel}.

\section{Systematic uncertainties}
\label{systematics}

The sources of systematic uncertainties considered in this analysis are listed below.  To determine the uncertainties in signal and background, each kinematic quantity listed is varied individually according to its uncertainty and the final event yields are re-measured to determine the variation in the predicted number of background and signal events.

Jet energy scale and resolution uncertainties are estimated by assigning $\PT$- and $\eta$-dependent uncertainties in jet energy corrections as discussed in Ref.~\cite{CMS-PAPERS-JME-10-011}, and varying the jet $\PT$ according to the magnitude of that uncertainty.  The uncertainty in the jet energy resolution is assessed by modifying the $\PT$ difference between the particle level and reconstructed jets by an $\eta$-dependent value between 5\% and 30\% for most jets~\cite{CMS-PAPERS-JME-10-011}.

Uncertainties in the charged-lepton momentum scale and resolution also introduce uncertainties in the final event acceptance.  An energy scale uncertainty of 0.6\% in the ECAL barrel and 1.5\% in the ECAL endcap is assigned to electrons~\cite{EleUncertainties}, and an uncertainty of 10\% in both the ECAL barrel and endcap is applied to the electron energy resolution~\cite{EleUncertainties}.  There is an uncertainty of 0.6\% per electron in reconstruction, identification, and isolation requirements.  For muons, a $\PT$-dependent scale uncertainty of  $5\% \, (\PT/1 \TeV)$ is applied, as well as a 1--4\% $\PT$-dependent resolution uncertainty~\cite{Chatrchyan:2012xi}.  In the case of momentum scale uncertainties the momentum is directly varied, and in the case of momentum resolution uncertainties the lepton momentum is subjected to a Gaussian random smearing within the uncertainty.  A 2\% per muon uncertainty in reconstruction, identification, and isolation requirements, as well as a 1\% muon HLT efficiency uncertainty, are assumed as well.

Other important sources of systematic uncertainty are related to the modeling of the backgrounds in the simulation.  The uncertainty in the $\cPZ/\gamma^{*}$ $+$ jets background shape  is determined by using simulated samples with renormalization and factorization scales and matrix-element parton-shower matching thresholds varied by a factor of two up and down.  The scale factors for the normalization of the $\cPZ/\gamma^{*}+$ jets background are assigned an uncertainty of 0.6\% in both channels, and the normalization of the \ttbar background is assigned an uncertainty of 0.5\% in both channels, based on the statistical uncertainties measured in the studies described in Section~\ref{backgrounds}.  An additional uncertainty of 4\% is applied to the $\ttbar$ background normalization in the $\Pgm \Pgm j$ channel to account for possible signal contamination from first generation LQs in the control sample (the contamination is extremely small in the other channel because of the suppressed second generation signal).  An uncertainty on $\cPZ/\gamma^{*}+$ jets background from the $\PT(\ell\ell)$ scale factors is assessed by taking the weighted average of the uncertainties from each $\PT(\ell\ell)$ bin.  The estimate of the QCD multijet background from data has an uncertainty of 15\%.

An uncertainty in the modeling of pileup in simulation is determined by varying the number of simulated pileup interactions up and down by 6\%~\cite{pileupuncert}, and an uncertainty of 2.6\% on the measured integrated luminosity is applied~\cite{CMS-PAS-LUM-13-001}.

Uncertainties in the signal acceptance, the background acceptance, and the cross sections, due to the PDF choice of 4--10\% for signal and 3--9\% for background are applied, following the PDF4LHC recommendations described in Refs.~\cite{PDF4LHC,Alekhin:2011sk}.

Finally, a statistical uncertainty associated with the size of the simulated sample is included for both background and signal.

The systematic uncertainties are listed in Table~\ref{tab:systs}, together with their effects on signal and background yields, corresponding to the final selection values optimized for $M_{\text{LQ}} = 600$ \GeV.  The PDF uncertainty is larger in the $\Pgm \Pgm j$ channel because of the large uncertainty associated with the $\PQs$-quark PDF.

\begin{table}
\topcaption{Systematic uncertainties (in \%) and their effects on total signal (S) and background (B) in both channels for $M_{\text{LQ}} = 600$ \GeV final selection.}
\centering
\begin{scotch}{lcccc}
{Systematic} & \multicolumn{2}{c}{$\Pe \Pe j$} & \multicolumn{2}{c}{$\Pgm \Pgm j$}\\
{uncertainty} & {S(\%)} & {B(\%)} & {S(\%)} & {B(\%)}\\
\hline
{Jet energy scale} & {0.3} & {1.0} & {0.7} & {1.4}\\
{Jet energy resolution}  & {0.1} & {0.3}  & {0.3} & {0.4}\\
{Electron energy scale}  & {0.2} & {2.1} & {\NA} & {\NA}\\
{Electron energy resolution}  & {0.1} & {0.6} & {\NA} & {\NA}\\
{Muon energy scale}  & {\NA} & {\NA} & {2.4} & {3.7}\\
{Muon energy resolution}  & {\NA} & {\NA} & {0.2} & {1.1}\\
{Electron reco/ID/iso}  & {1.2} & {0.1} & {\NA} & {\NA}\\
{Muon reco/ID/iso}  & {\NA} & {\NA} & {2.0} & {0.1}\\
{Trigger}  & {\NA} & {\NA}  & {1.0} & {0.1}\\
{QCD normalization}  & {\NA} & {0.0} & {\NA} & {0.1}\\
{\ttbar normalization}  & {\NA} & {0.2}  & {\NA} & {1.1}\\
{$\cPZ/\gamma^* +{\rm jets}$ normalization}  & {\NA} & {0.3} & {\NA} & {0.3}\\
{$\cPZ/\gamma^* +{\rm jets}$ shape}  & {\NA} & {5.2} & {\NA} & {5.6}\\
{$\cPZ/\gamma^* +{\rm jets}$ $\PT(\ell\ell)$ scale factor}  & {\NA} & {2.6} & {\NA} & {3.0}\\
{PDF}  & {3.5} & {3.0}  & {3.0} & {2.8}\\
{Pileup}  & {2.5} & {0.6} & {2.8} & {1.9}\\
{Integrated luminosity} & {2.6} & {0.3} & {2.6} & {0.2}\\
{Statistical uncertainty} & {1.3} & {3.5} & {1.4} & {4.3}\\
\hline
{Total}  & {5.3} & {8.1}  & {6.05} & {8.1}\\
\end{scotch}
\label{tab:systs}
\end{table}

\section{Results}
\label{results}
The observed data are consistent with the no-signal hypothesis.  We set an upper limit on the leptoquark cross section by using the \CLS modified frequentist method~\cite{Read:2002hq,Junk:1999kv} with the final event yields.  A log-normal probability function is used to model the systematic uncertainties, whereas statistical uncertainties are described with gamma distributions with widths determined according to the number of events simulated or measured in data control regions.

To isolate the limits for resonant LQ production, we apply the resonant requirements at the generator level on both the lepton+jet mass, $M(\ell,j)> (0.67 \text{ or } 0.75 )\, M_{\text{LQ}}$ (for the first- or second- generation LQs, respectively), and on the dilepton mass, $M_{\ell \ell}> 110$\GeV.  These requirements make the limits extracted from data more conservative and are discussed in Section~\ref{selection}.  A resonant cross section $\sigma_{\text{res}}$ is computed with respect to those requirements.  Limits are then computed with the reduced sample of simulated signal events and compared to $\sigma_{\text{res}}$.

\begin{figure}[!htbp]
       \centering
       {\includegraphics[width=.45\textwidth]{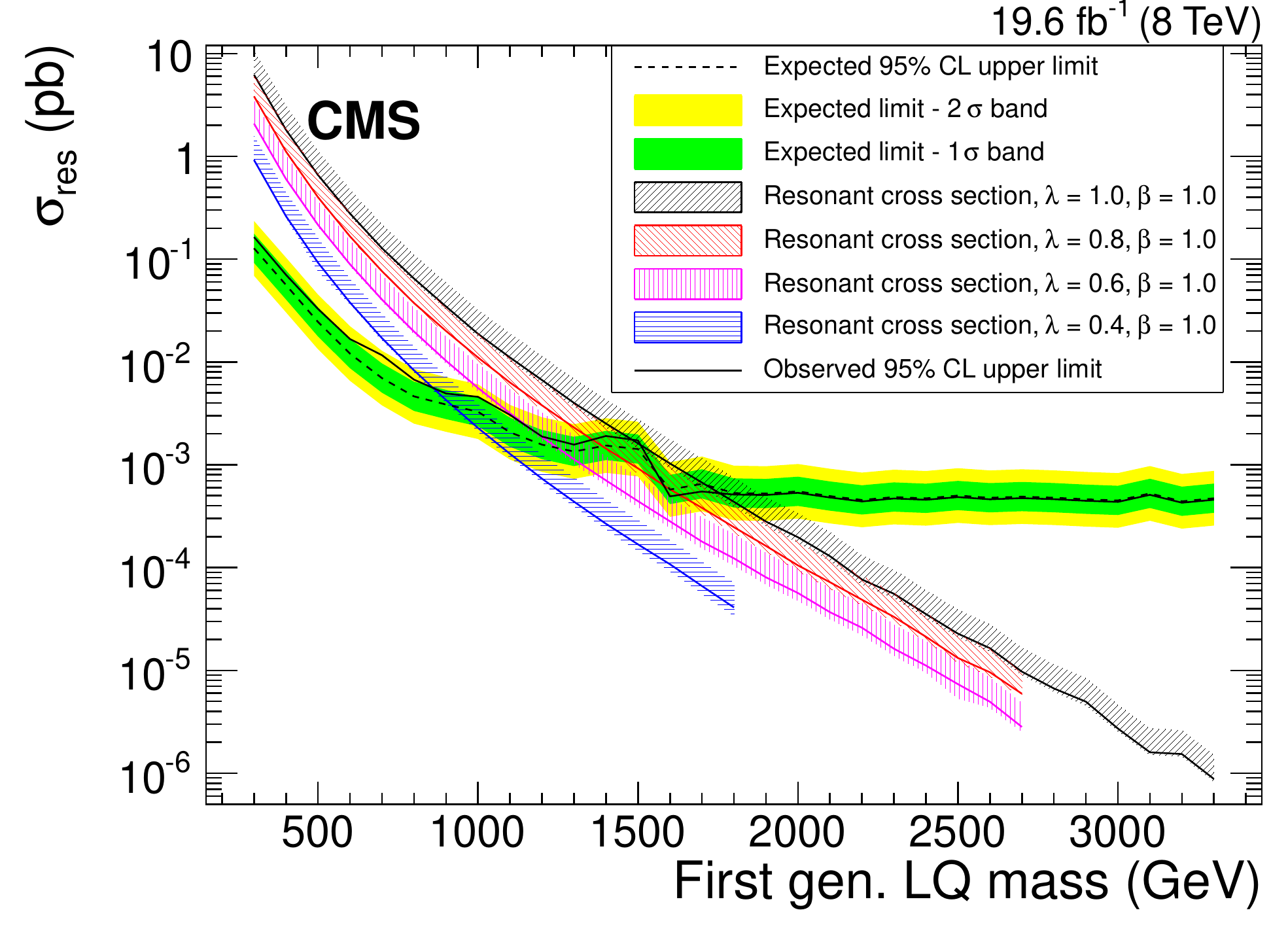}}
       {\includegraphics[width=.45\textwidth]{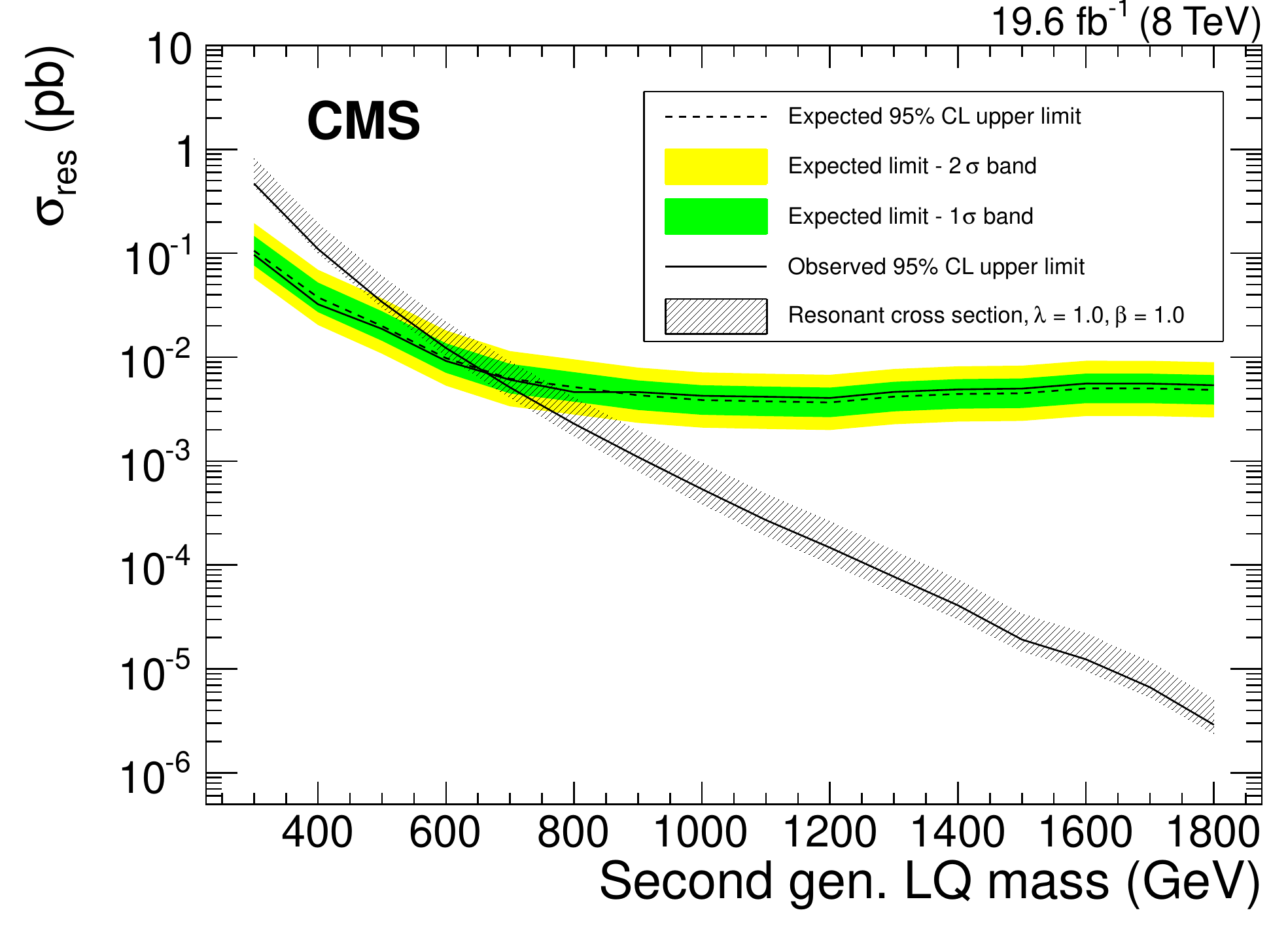}}
       \caption{Expected and observed upper limits at 95\%~\CL on first and second generation leptoquark single production resonant cross section as a function of the leptoquark mass.  First generation limits are shown on the \cmsLeft\ plot with a resonant region of $M_{\ell j} > 0.66 \, M_{\text{LQ}}$, $M_{\ell \ell}> 110$\GeV and second generation limits are shown on the \cmsRight\ plot with a resonant region of $M_{\ell j} > 0.75 \, M_{\text{LQ}}$, $M_{\ell \ell}> 110$\GeV.  The uncertainty bands on the observed limit represent the 68\% and 95\% confidence intervals.  The uncertainty band on the theoretical cross section includes uncertainties due to PDF variation and the k factor.\label{figapp:combinedlims}}
\end{figure}

The 95\% confidence level (\CL) upper limits on $\sigma_{\text{res}} \, \beta$ as a function of leptoquark mass are shown in Fig.~\ref{figapp:combinedlims} together with the resonant cross section predictions for the scalar leptoquark single production cross section.  The uncertainty band on the theoretical cross section prediction corresponds to uncertainties in the total cross section due to PDF variations with an additional $+70\%$ uncertainty, because of the k factor from NLO corrections~\cite{Hammett:2015sea}.  The observed limits are listed in Tables \ref{tab:optimizedcuts1p0} and \ref{tab:optimizedcutscmu} in the appendix.

By comparing the observed upper limit with the theoretical production cross-section times branching fraction, we exclude single leptoquark production at 95\% ~\CL for LQ masses below the values given in Table~\ref{figapp:limitstable}.

\begin{table}[!htbp]
\topcaption{95\% ~\CL lower limits on scalar LQ masses ($\beta = 1.0$).}
\centering
\begin{scotch}{rc}
LQ generation, coupling & Excluded mass (\GeVns) \\
\hline
{First gen., $\lambda=0.4$} & {   860} \\
{First gen., $\lambda=0.6$} & {1175} \\
{First gen., $\lambda=0.8$} & {1355} \\
{First gen., $\lambda=1.0$} & {1755} \\
{Second gen., $\lambda=1.0$} & {     660} \\
\end{scotch}
\label{figapp:limitstable}
\end{table}

Limits on single production of the $S^{R}_{0}$ type LQ from the H1 collaboration exclude LQ production up to 500\GeV ($\lambda=1.0$) and up to 350\GeV ($\lambda=0.6$)~\cite{Collaboration:2011qaa}.

\section{Summary}
\label{conclusion}

A search has been performed for the single production of first- and second- generation scalar leptoquarks in final states with two electrons and a jet or two muons and a jet using a data set of proton-proton collisions at 8 \TeV corresponding to an integrated luminosity of 19.6 fb$^{-1}$.  The selection criteria are optimized for each leptoquark signal mass hypothesis.  The number of observed candidates for each mass hypothesis agrees with the number of expected standard model background events.  Single production of first-(second-) generation leptoquarks with a coupling of 1.0 is excluded at 95\% confidence level for masses below 1755(660)\GeV.  These are the most stringent limits to date for single production.  The first-generation limits for couplings greater than 0.6 are stronger than those from pair production and are the most stringent overall limits on leptoquark production in the first generation to date.

\section*{Acknowledgments}

\hyphenation{Bundes-ministerium Forschungs-gemeinschaft Forschungs-zentren} We congratulate our colleagues in the CERN accelerator departments for the excellent performance of the LHC and thank the technical and administrative staffs at CERN and at other CMS institutes for their contributions to the success of the CMS effort. In addition, we gratefully acknowledge the computing centers and personnel of the Worldwide LHC Computing Grid for delivering so effectively the computing infrastructure essential to our analyses. Finally, we acknowledge the enduring support for the construction and operation of the LHC and the CMS detector provided by the following funding agencies: the Austrian Federal Ministry of Science, Research and Economy and the Austrian Science Fund; the Belgian Fonds de la Recherche Scientifique, and Fonds voor Wetenschappelijk Onderzoek; the Brazilian Funding Agencies (CNPq, CAPES, FAPERJ, and FAPESP); the Bulgarian Ministry of Education and Science; CERN; the Chinese Academy of Sciences, Ministry of Science and Technology, and National Natural Science Foundation of China; the Colombian Funding Agency (COLCIENCIAS); the Croatian Ministry of Science, Education and Sport, and the Croatian Science Foundation; the Research Promotion Foundation, Cyprus; the Ministry of Education and Research, Estonian Research Council via IUT23-4 and IUT23-6 and European Regional Development Fund, Estonia; the Academy of Finland, Finnish Ministry of Education and Culture, and Helsinki Institute of Physics; the Institut National de Physique Nucl\'eaire et de Physique des Particules~/~CNRS, and Commissariat \`a l'\'Energie Atomique et aux \'Energies Alternatives~/~CEA, France; the Bundesministerium f\"ur Bildung und Forschung, Deutsche Forschungsgemeinschaft, and Helmholtz-Gemeinschaft Deutscher Forschungszentren, Germany; the General Secretariat for Research and Technology, Greece; the National Scientific Research Foundation, and National Innovation Office, Hungary; the Department of Atomic Energy and the Department of Science and Technology, India; the Institute for Studies in Theoretical Physics and Mathematics, Iran; the Science Foundation, Ireland; the Istituto Nazionale di Fisica Nucleare, Italy; the Ministry of Science, ICT and Future Planning, and National Research Foundation (NRF), Republic of Korea; the Lithuanian Academy of Sciences; the Ministry of Education, and University of Malaya (Malaysia); the Mexican Funding Agencies (CINVESTAV, CONACYT, SEP, and UASLP-FAI); the Ministry of Business, Innovation and Employment, New Zealand; the Pakistan Atomic Energy Commission; the Ministry of Science and Higher Education and the National Science Centre, Poland; the Funda\c{c}\~ao para a Ci\^encia e a Tecnologia, Portugal; JINR, Dubna; the Ministry of Education and Science of the Russian Federation, the Federal Agency of Atomic Energy of the Russian Federation, Russian Academy of Sciences, and the Russian Foundation for Basic Research; the Ministry of Education, Science and Technological Development of Serbia; the Secretar\'{\i}a de Estado de Investigaci\'on, Desarrollo e Innovaci\'on and Programa Consolider-Ingenio 2010, Spain; the Swiss Funding Agencies (ETH Board, ETH Zurich, PSI, SNF, UniZH, Canton Zurich, and SER); the Ministry of Science and Technology, Taipei; the Thailand Center of Excellence in Physics, the Institute for the Promotion of Teaching Science and Technology of Thailand, Special Task Force for Activating Research and the National Science and Technology Development Agency of Thailand; the Scientific and Technical Research Council of Turkey, and Turkish Atomic Energy Authority; the National Academy of Sciences of Ukraine, and State Fund for Fundamental Researches, Ukraine; the Science and Technology Facilities Council, UK; the US Department of Energy, and the US National Science Foundation.

Individuals have received support from the Marie-Curie program and the European Research Council and EPLANET (European Union); the Leventis Foundation; the A. P. Sloan Foundation; the Alexander von Humboldt Foundation; the Belgian Federal Science Policy Office; the Fonds pour la Formation \`a la Recherche dans l'Industrie et dans l'Agriculture (FRIA-Belgium); the Agentschap voor Innovatie door Wetenschap en Technologie (IWT-Belgium); the Ministry of Education, Youth and Sports (MEYS) of the Czech Republic; the Council of Science and Industrial Research, India; the HOMING PLUS program of the Foundation for Polish Science, cofinanced from European Union, Regional Development Fund; the OPUS program of the National Science Center (Poland); the Compagnia di San Paolo (Torino); the Consorzio per la Fisica (Trieste); MIUR project 20108T4XTM (Italy); the Thalis and Aristeia programs cofinanced by EU-ESF and the Greek NSRF; the National Priorities Research Program by Qatar National Research Fund; the Rachadapisek Sompot Fund for Postdoctoral Fellowship, Chulalongkorn University (Thailand); and the Welch Foundation, contract C-1845.

\bibliography{auto_generated}
\numberwithin{table}{section}
\appendix
\label{appendix}

\section{Signal cross sections}

This section contains a table of first- and second-generation LQ cross sections, computed at LO in {{\textsc{CalcHEP}}\xspace} and scaled for the resonant selection (Table~\ref{tab:sigxsectionscombined}).

\begin{table*}[!htbp]
\topcaption{Signal cross sections calculated at LO in {{\textsc{CalcHEP}}\xspace}.  Resonant cross sections scaled by the acceptance of the selections described in Section~\ref{selection} are listed under each corresponding LO cross section.}
\centering
  \begin{scotch}{r.....}
\multicolumn{1}{c}{$M_{\text{LQ}}$}  & \multicolumn{1}{c}{First gen.,} &\multicolumn{1}{c}{First gen.,} &\multicolumn{1}{c}{First gen.,} &\multicolumn{1}{c}{First gen.,} &\multicolumn{1}{c}{Second gen.,} \\
\multicolumn{1}{c}{($\GeVns$)} & \multicolumn{1}{c}{$\lambda=0.4$} & \multicolumn{1}{c}{$\lambda=0.6$} &\multicolumn{1}{c}{$\lambda=0.8$} & \multicolumn{1}{c}{$\lambda=1.0$} & \multicolumn{1}{c}{$\lambda=1.0$} \\
& \multicolumn{1}{c}{(pb)} & \multicolumn{1}{c}{(pb)} & \multicolumn{1}{c}{(pb)} & \multicolumn{1}{c}{(pb)} & \multicolumn{1}{c}{(pb)} \\
\hline
300&1.04&2.39&4.38&7.12&0.579\\
&0.921&2.08&3.83&6.21&0.468\\
400&0.291&0.675&1.25&2.06&0.139\\
&0.261&0.601&1.11&1.81&0.11\\
500&0.102&0.239&0.451&0.755&0.0446\\
&0.0924&0.215&0.4&0.658&0.034\\
600&0.0413&0.0984&0.189&0.322&0.0176\\
&0.0378&0.0891&0.166&0.278&0.0122\\
700&0.0186&0.0451&0.088&0.154&0.00807\\
&0.017&0.0404&0.0763&0.128&0.00511\\
800&0.00904&0.0223&0.0446&0.0797&0.00418\\
&0.00829&0.0198&0.0374&0.0647&0.00229\\
900&0.00467&0.0118&0.0242&0.0443&0.00237\\
&0.00427&0.0103&0.02&0.0346&0.00109\\
1000&0.00254&0.00657&0.0139&0.0261&0.00145\\
&0.00228&0.00559&0.0111&0.0188&0.000537\\
1200&0.00084&0.00234&0.00526&0.0104&0.00064\\
&0.000733&0.00186&0.00378&0.00667&0.000147\\
1400&0.00032&0.00097&0.00233&0.00485&0.00033\\
&0.000267&0.000705&0.00144&0.00252&\multicolumn{1}{r}{4.09e-05}\\
1600&0.00014&0.00045&0.00117&0.00255&0.00019\\
&0.000108&0.000282&0.000577&0.00103&\multicolumn{1}{r}{1.24e-05}\\
1800&\multicolumn{1}{r}{6e-05}&0.00024&0.00065&0.00147&0.00011\\
&\multicolumn{1}{r}{4.1e-05}&0.000123&0.000247&0.000436&\multicolumn{1}{r}{2.9e-06}\\
2000&&0.00014&0.00039&0.00092&\\
&&\multicolumn{1}{r}{5.66e-05}&0.000105&0.000197&\\
2500&&\multicolumn{1}{r}{5e-05}&0.00014&0.00035&\\
&&\multicolumn{1}{r}{7.35e-06}&\multicolumn{1}{r}{1.32e-05}&\multicolumn{1}{r}{2.28e-05}&\\
3000&&&&0.00016&\\
&&&&\multicolumn{1}{r}{2.72e-06}&\\
3300&&&&0.00011&\\
&&&&\multicolumn{1}{r}{8.79e-07}&\\
  \end{scotch}
\label{tab:sigxsectionscombined}
\end{table*}

\section{Final selection}

This section contains the reference tables for the final selection criteria and the corresponding observed limits for the $\Pe \Pe j$ and the $\Pgm \Pgm j$ channels.
\begin{table*}[!htbp]
\topcaption{The $\Pe \Pe j$ channel threshold values for $\ST$, $M_{\Pe j}$, and $M_{\Pe j, \text{gen}}$ \vs LQ mass (for all couplings), and the corresponding observed limits.}
       \centering
\begin{scotch}{cccc.}
{$M_{\text{LQ}}$} & {$\ST$ threshold} & {$M_{\Pe j}$ threshold} & {$M_{\Pe j, \text{gen}}$ threshold} & \multicolumn{1}{c}{Observed limit on $\sigma_{\text{res}}$}\\
{($\GeVns$)} & {($\GeVns$)} & {($\GeVns$)} & {($\GeVns$)} & \multicolumn{1}{c}{(pb)}\\
\hline
300&250&200&200&0.16\\
400&320&300&266&0.07\\
500&400&400&333&0.033\\
600&480&500&400&0.017\\
700&560&600&466&0.012\\
800&640&700&533&0.0067\\
900&720&800&600&0.0049\\
1000&800&900&666&0.0046\\
1200&900&1100&800&0.0019\\
1400&900&1300&933&0.0019\\
1600&900&1500&1066&0.00049\\
1800&900&1700&1200&0.00051\\
2000&900&1900&1333&0.00053\\
2500&900&1900&1666&0.00048\\
3000&900&1900&2000&0.00044\\
3300&900&1900&2200&0.00046\\
\end{scotch}
\label{tab:optimizedcuts1p0}
\end{table*}

\begin{table*}[!htbp]
\topcaption{The $\Pgm \Pgm j$ channel threshold values for $\ST$, $M_{\Pgm j}$, and $M_{\Pgm j, \text{gen}}$ \vs LQ mass, and the corresponding observed limits.}
       \centering
\begin{scotch}{cccc.}
{$M_{\text{LQ}}$} & {$\ST$ threshold} & {$M_{\Pgm j}$ threshold}& {$M_{\Pgm j, \text{gen}}$ threshold}& \multicolumn{1}{c}{Observed limit on $\sigma_{\text{res}}$}\\
{($\GeVns$)} & {($\GeVns$)} & {($\GeVns$)} & {($\GeVns$)} & \multicolumn{1}{c}{(pb)}\\
\hline
300&300&200&225&0.096\\
400&400&300&300&0.032\\
500&500&400&375&0.019\\
600&600&500&450&0.0092\\
700&700&600&525&0.0061\\
800&800&700&600&0.0046\\
900&900&800&675&0.0046\\
1000&1000&800&750&0.0042\\
1200&1000&800&900&0.004\\
1400&1000&800&1050&0.0049\\
1600&1000&800&1200&0.0056\\
1800&1000&800&1350&0.0054\\
\end{scotch}
\label{tab:optimizedcutscmu}
\end{table*}

\section{Event yields}

This section contains tables of data, background, and signal yields after the final selection.   Event counts vary between the two channels due to differences in the optimized thresholds for $\ST$ and $M_{\ell j}$ as well as differences in the electron and muon efficiencies.  The first listed uncertainty is statistical, the second is systematic; in cases where only one uncertainty is listed it is statistical.

\begin{table*}[!htbp]
\topcaption{Data and background yields after final selection for the $\Pe \Pe j$ channel for first-generation LQs, shown with statistical and systematic uncertainties.  ``Other backgrounds'' refers to diboson+jets, $\PW $+ ${\rm jets}$, single-top quark, and QCD.  The values do not change above 2000\GeV.}
       \centering
\begin{scotch}{cccccc}
{$M_{\text{LQ}} (\GeVns)$} & {Data} & {Total background} & {$\cPZ/\gamma^*$ + ${\rm jets}$} & {$\ttbar$} & { Other backgrounds }\\
\hline
{$300$} & {$3007$} & {$2830\pm 40\pm 170$} & {$1362\pm 19$} & {$1238\pm 27$} & {$230\pm 15$}\\
{$400$} & {$1766$} & {$1660\pm 30\pm 110$} & {$873\pm 15$} & {$637\pm 19$} & {$151\pm 12$}\\
{$500$} & {$807$} & {$736\pm 18\pm 49$} & {$409.8\pm 9.6$} & {$251\pm 12$} & {$75.6\pm 8.6$}\\
{$600$} & {$370$} & {$329\pm 12\pm 24$} & {$192.9\pm 6.3$} & {$102.7\pm 7.9$} & {$33.3\pm 5.8$}\\
{$700$} & {$186$} & {$149\pm 8\pm 12$} & {$91.6\pm 4.1$} & {$40.9\pm 4.9$} & {$16.7\pm 4.2$}\\
{$800$} & {$91$} & {$73.7\pm 5.6\pm 7.0$} & {$46.3\pm 2.8$} & {$21.1\pm 3.5$} & {$6.3\pm 3.3$}\\
{$900$} & {$46$} & {$36.9\pm 3.4\pm 6.6$} & {$23.9\pm 1.9$} & {$7.6\pm 2.1$} & {$5.5\pm 1.9$}\\
{$1000$} & {$28$} & {$18.3\pm 2.5\pm 4.8$} & {$11.7\pm 1.3$} & {$3.7\pm 1.5$} & {$2.9\pm 1.5$}\\
{\rule{0pt}{1ex}$1200$} & {$7$} & {$5.2\pm 1.6\pm 1.8$} & {$3.17\pm 0.61$} & {$0.39^{+0.53}_{-0.39}$} & {$1.6\pm 1.3$}\\
{\rule{0pt}{3ex}$1400$} & {$4$} & {$1.8\pm 1.3\pm 1.5$} & {$1.0\pm 0.31$} & {$0.0^{+0.41}_{-0.0}$} & {$0.8^{+1.2}_{-0.8}$}\\
{\rule{0pt}{3ex}$1600$} & {$0$} & {$0.2^{+1.2}_{-0.2}{}^{+0.4}_{-0.2}$} & {$0.17\pm 0.12$} & {$0.0^{+0.41}_{-0.0}$} & {$0.1^{+1.2}_{-0.1}$}\\
{\rule{0pt}{3ex}$1800$} & {$0$} & {$0.0^{+1.3}_{-0.0}\pm 0.0$} & {$0.0^{+0.22}_{-0.0}$} & {$0.0^{+0.41}_{-0.0}$} & {$0.0^{+1.2}_{-0.0}$}\\
{\rule{0pt}{3ex}$2000$} & {$0$} & {$0.0^{+1.3}_{-0.0}\pm 0.0$} & {$0.0^{+0.22}_{-0.0}$} & {$0.0^{+0.41}_{-0.0}$} & {$0.0^{+1.2}_{-0.0}$}\\
\end{scotch}
\label{tab:eenumbers1p0}
\end{table*}

\begin{table*}[!htbp]
\topcaption{Signal yields after final selection in the $\Pe \Pe j$ channel for first-generation LQs shown with statistical and systematic uncertainties, for different values of $\lambda$ and for $\beta = 1.0$.}
       \centering
\hspace*{-0.5cm}\begin{scotch}{ccccc}
{$M_{\text{LQ}}$} & {$\lambda=0.4$} & {$\lambda=0.6$} & {$\lambda=0.8$} & {$\lambda=1.0$} \\
{($\GeVns$)} &  &  &  & \\
\hline
300&{$3540\pm 60\pm 200$}&{$7880\pm 130\pm 420$}&{$14390\pm 240\pm 820$}&{$22600\pm 400\pm 1200$}\\
400&{$1577\pm 22\pm 85$}&{$3600\pm 50\pm 190$}&{$6330\pm 80\pm 340$}&{$9990\pm 150\pm 530$}\\
500&{$670\pm 10\pm 160$}&{$1504\pm 18\pm 85$}&{$2670\pm 30\pm 140$}&{$4270\pm 60\pm 210$}\\
600&{$289\pm 3\pm 18$}&{$666\pm 8\pm 33$}&{$1188\pm 14\pm 76$}&{$1920\pm 30\pm 100$}\\
700&{$138.1\pm 1.6\pm 6.2$}&{$320\pm 4\pm 15$}&{$559\pm 7\pm 27$}&{$885\pm 12\pm 41$}\\
800&{$67.8\pm 0.8\pm 3.3$}&{$158.2\pm 1.8\pm 6.5$}&{$275\pm 3\pm 12$}&{$446\pm 6\pm 19$}\\
900&{$35.9\pm 0.4\pm 1.4$}&{$82.5\pm 0.9\pm 3.3$}&{$145.7\pm 1.8\pm 5.6$}&{$231\pm 3\pm 11$}\\
1000&{$19.26\pm 0.22\pm 0.88$}&{$43.6\pm 0.5\pm 1.8$}&{$77.9\pm 1.0\pm 3.1$}&{$118.3\pm 1.8\pm 4.6$}\\
1200&{$6.14\pm 0.07\pm 0.25$}&{$13.8\pm 0.2\pm 1.2$}&{$25.44\pm 0.35\pm 0.98$}&{$39.7\pm 0.6\pm 1.9$}\\
1400&{$2.2\pm 0.0\pm 0.2$}&{$5.07\pm 0.07\pm 0.28$}&{$9.13\pm 0.14\pm 0.58$}&{$13.78\pm 0.26\pm 0.88$}\\
1600&{$0.8\pm 0.0\pm 0.1$}&{$1.89\pm 0.03\pm 0.15$}&{$3.3\pm 0.06\pm 0.26$}&{$5.24\pm 0.12\pm 0.46$}\\
1800&{$0.29\pm 0.0\pm 0.03$}&{$0.76\pm 0.01\pm 0.08$}&{$1.31\pm 0.03\pm 0.13$}&{$2.02\pm 0.06\pm 0.24$}\\
2000&&{$0.31\pm 0.01\pm 0.04$}&{$0.497\pm 0.014\pm 0.071$}&{$0.81\pm 0.03\pm 0.12$}\\
2500&&{$0.039\pm 0.001\pm 0.032$}&{$0.064\pm 0.003\pm 0.016$}&{$0.102\pm 0.006\pm 0.023$}\\
3000&&&&{$0.0134\pm 0.0015\pm 0.0029$}\\
3300&&&&{$0.004\pm 0.001\pm 0.001$}\\
\end{scotch}
\label{tab:eesignumbers1p0}
\end{table*}

\begin{table*}[!htbp]
\topcaption{Data, signal, and background yields after final selection in the $\Pgm \Pgm j$ channel shown with statistical and total systematic uncertainties, for $\lambda = 1.0$ and $\beta = 1.0$.  ``Other backgrounds'' refers to diboson+jets, $\PW $+ ${\rm jets}$, single-top quark, and QCD.}
       \centering
\hspace*{-1.5cm}\begin{scotch}{ccccccc}
{$M_{\text{LQ}} (\GeVns)$} & {Signal} & {Data} & {Total background} & {$\cPZ/\gamma^*$ + ${\rm jets}$} & {$\ttbar$} & { Other backgrounds}\\
\hline
{$300$} & {$2130\pm 30\pm 290$}& {$3036$} & {$3120\pm 40\pm 370$} & {$1541\pm 20$} & {$1362\pm 32$} & {$214\pm 15$}\\
{$400$} & {$721\pm 9\pm 91$}& {$1371$} & {$1440\pm 30\pm 170$} & {$774\pm 14$} & {$548\pm 21$} & {$118\pm 11$}\\
{$500$} & {$228\pm 3\pm 27$}& {$558$} & {$577\pm 17\pm 75$} & {$340.7\pm 8.6$} & {$182\pm 12$} & {$54.3\pm 8.1$}\\
{$600$} & {$77.1\pm 1.1\pm 9.5$}& {$238$} & {$246\pm 10\pm 32$} & {$155.6\pm 5.6$} & {$73.8\pm 7.7$} & {$16.4\pm 4.3$}\\
{$700$} & {$28.0\pm 0.5\pm 3.7$}& {$100$} & {$102\pm 6\pm 14$} & {$70.1\pm 3.5$} & {$22.3\pm 4.3$} & {$9.5\pm 2.7$}\\
{$800$} & {$10.7\pm 0.2\pm 1.6$}& {$48$} & {$52.3\pm 4.7\pm 7.6$} & {$32.3\pm 2.3$} & {$12.3\pm 3.2$} & {$7.7\pm 2.6$}\\
{$900$} & {$4.67\pm 0.1\pm 0.84$}& {$27$} & {$25.7\pm 3.5\pm 4.6$} & {$14.9\pm 1.5$} & {$4.8\pm 2.0$} & {$5.9\pm 2.5$}\\
{$1000$} & {$2.1\pm 0.05\pm 0.46$}& {$17$} & {$15.5\pm 3.0\pm 3.3$} & {$7.6\pm 1.1$} & {$2.6\pm 1.5$} & {$5.3\pm 2.4$}\\
{$1200$} & {$0.7\pm 0.02\pm 0.22$}& {$17$} & {$15.5\pm 3.0\pm 3.3$} & {$7.6\pm 1.1$} & {$2.6\pm 1.5$} & {$5.3\pm 2.4$}\\
{$1400$} & {$0.195\pm 0.008\pm 0.088$}& {$17$} & {$15.5\pm 3.0\pm 3.3$} & {$7.6\pm 1.1$} & {$2.6\pm 1.5$} & {$5.3\pm 2.4$}\\
{$1600$} & {$0.06\pm 0.003\pm 0.032$}& {$17$} & {$15.5\pm 3.0\pm 3.3$} & {$7.6\pm 1.1$} & {$2.6\pm 1.5$} & {$5.3\pm 2.4$}\\
{$1800$} & {$0.0135\pm 0.0012\pm 0.0066$}& {$17$} & {$15.5\pm 3.0\pm 3.3$} & {$7.6\pm 1.1$} & {$2.6\pm 1.5$} & {$5.3\pm 2.4$}\\
\end{scotch}
\label{tab:mumunumbers}
\end{table*}

\cleardoublepage \section{The CMS Collaboration \label{app:collab}}\begin{sloppypar}\hyphenpenalty=5000\widowpenalty=500\clubpenalty=5000\textbf{Yerevan Physics Institute,  Yerevan,  Armenia}\\*[0pt]
V.~Khachatryan, A.M.~Sirunyan, A.~Tumasyan
\vskip\cmsinstskip
\textbf{Institut f\"{u}r Hochenergiephysik der OeAW,  Wien,  Austria}\\*[0pt]
W.~Adam, E.~Asilar, T.~Bergauer, J.~Brandstetter, E.~Brondolin, M.~Dragicevic, J.~Er\"{o}, M.~Flechl, M.~Friedl, R.~Fr\"{u}hwirth\cmsAuthorMark{1}, V.M.~Ghete, C.~Hartl, N.~H\"{o}rmann, J.~Hrubec, M.~Jeitler\cmsAuthorMark{1}, V.~Kn\"{u}nz, A.~K\"{o}nig, M.~Krammer\cmsAuthorMark{1}, I.~Kr\"{a}tschmer, D.~Liko, T.~Matsushita, I.~Mikulec, D.~Rabady\cmsAuthorMark{2}, B.~Rahbaran, H.~Rohringer, J.~Schieck\cmsAuthorMark{1}, R.~Sch\"{o}fbeck, J.~Strauss, W.~Treberer-Treberspurg, W.~Waltenberger, C.-E.~Wulz\cmsAuthorMark{1}
% ...
\vskip\cmsinstskip
% ...
1:~~Also at Vienna University of Technology, Vienna, Austria\\
2:~~Also at CERN, European Organization for Nuclear Research, Geneva, Switzerland\\
% ...

\end{sloppypar}
\end{document}